\documentclass{jfm}
\usepackage[version=3]{mhchem}
\usepackage[subrefformat=parens,labelformat=parens]{subfig} 
\usepackage{placeins} 
\usepackage{xcolor} 
\usepackage[normalem]{ulem}
\usepackage{siunitx}
\newcommand{\vect}[1]{\ensuremath{\boldsymbol{#1}}}

\newcommand{\dif}{\mathop{}\mathopen{}\mathrm{d}}
\newcommand{\ez}{\vect{e}_z}
\newcommand{\er}{\vect{e}_r}
\newcommand{\Prandtl}{\ensuremath{\Pran}} 
\newcommand{\Pm}{\ensuremath{Pm}}
\newcommand{\Rm}{\ensuremath{Rm}}
\newcommand{\Ra}{\ensuremath{{Ra}}}
\newcommand{\Rac}{\ensuremath{{Ra_\text{c}}}}
\newcommand{\Ekman}{\ensuremath{E}}
\newcommand{\Di}{\ensuremath{Di}}
\newcommand{\Nrho}{\ensuremath{N_\varrho}}
\newcommand{\npol}{\ensuremath{n}}
\newcommand{\aspectratio}{\ensuremath{\chi}}
\newcommand{\Ek}{\ensuremath{E_k}}
\newcommand{\Eb}{\ensuremath{E_b}}
\newcommand{\EkS}{\ensuremath{E_k^S}}
\newcommand{\EkA}{\ensuremath{E_k^A}}
\newcommand{\EbS}{\ensuremath{E_b^S}}
\newcommand{\EbA}{\ensuremath{E_b^A}}
\newcommand{\Ez}{\ensuremath{E_{Z}}}
\newcommand{\Tmod}{\ensuremath{T_\text{mod}}}
\newlength{\onefig}
\newlength{\threefig}
\begin{document}

\newtheorem{lemma}{Lemma}
\newtheorem{corollary}{Corollary}

\shorttitle{Convective dynamos: symmetries and modulation} 
\shortauthor{R.~Raynaud and S.~M.~Tobias} 

\title{Convective dynamo action in a spherical shell: symmetries and modulation}

\author
 {
 Rapha\"{e}l Raynaud\aff{1}\aff{2}
  \corresp{\email{raphael.raynaud@ipm.ir}}
  \and
  Steven M.~Tobias\aff{3}
  }

\affiliation
{
\aff{1} 
School of Astronomy, Institute for Research in
Fundamental Sciences (IPM), 19395-5531, Tehran, Iran
\aff{2} 
LERMA, Observatoire de Paris, PSL Research University, CNRS, Sorbonne
Universit\'{e}s, UPMC Univ. Paris 06, \'{E}cole normale
sup\'{e}rieure, F-75005, Paris, France
\aff{3} Department of Applied
Mathematics, University of Leeds, Leeds LS2 9JT, UK}

\maketitle

\begin{abstract}
We consider dynamo action driven by three-dimensional rotating
anelastic convection in a spherical shell. Motivated by the behaviour
of the solar dynamo, we examine the interaction of hydromagnetic modes
with different symmetries and demonstrate how complicated interactions
between convection, differential rotation and magnetic fields may
lead to modulation of the basic cycle. For some parameters, Type~1
modulation occurs by the transfer of energy between modes of different
symmetries with little change in the overall amplitude; for other
parameters, the modulation is of Type~2 where the amplitude is
significantly affected (leading to grand minima in activity) without
significant changes in symmetry.  Most importantly we identify the
presence of `supermodulation' in the solutions where the activity
switches chaotically between Type~1 and Type~2 modulation; this is
believed to be an important process in solar activity.
\end{abstract}

\section{Introduction}

The origin of magnetic activity in stellar interiors is a fundamental
problem of magnetohydrodynamics. The global solar magnetic field
oscillates with a mean period of twenty-two years (leading to an
eleven-year activity cycle) and is believed to be generated via a
dynamo acting (at least in part) deep within the Sun. The Sun's
magnetic field is largely dipolar; i.e. the mean azimuthal field that
leads to the formation of active regions is generally antisymmetric
about the equator. However, when this field is weak at the end of a
cycle, it takes on a more mixed character, with a quadrupole component
that becomes significant \citep{sokoloff1994}. Furthermore, direct
observations and proxy data demonstrate that the amplitude of the
solar cycle is modulated on longer time scales. There is indeed a
period of reduced activity between 1645 and 1715 ---~the Maunder
minimum~--- when the occurrence of sunspots was much reduced
\citep{eddy1976,usoskin2015}.  Analysis of the abundances of the
cosmogenic isotopes \ce{^{10}Be} in polar ice and \ce{^{14}C} in tree
rings reveals 27~grand minima in the past 11\,000~yr, separated by
aperiodic intervals of approximately 200~yr
\citep{usoskin2013,cracken2013}. A key observation for our
understanding of the processes leading to modulation is that as the
Sun emerged from the Maunder minimum, sunspots were largely restricted
to the southern hemisphere, showing that the magnetic field emerged
with a mixed character with both dipole and quadrupole components
\citep{sokoloff1994}.  Moreover, between 1750 and 1775, the solar
magnetic field took on a more quadrupolar character, with sunspots
appearing at the equator \citep{arlt2009}. There is now evidence from
the cosmogenic isotope records that the Sun switches on a long time
scale between strong modulation with clusters of deep grand minima and
weaker modulation, which can be associated with symmetry
breaking. This `supermodulation' is an example of chaotic (though
deterministic) modulational effects \citep{weisstobias2016}.  Evidence
for modulation in other stars arises from the long-term monitoring of
the CaII H+K flux of solar-type stars started by Wilson in 1968. The
so-called Mount Wilson Observatory survey provides a panel of
different stellar activities, in which \SI{60}{\percent} of stars
exhibit periodic variations, and \SI{25}{\percent} show
irregular or aperiodic variability
\citep{baliunas1998,olah2009}. Evidence for changes of symmetry in
young rapidly rotating stars is also now beginning to emerge
\citep{hlrk2016}.

Stellar magnetic fields are thought to be maintained against ohmic
dissipation by dynamo action through the flow of an electrically
conducting fluid \citep{moffattHK}.  Although it is known that
systematic activity can be generated through the interaction of
turbulent flows with rotation, shear and magnetic fields, no
satisfactory, self-consistent nonlinear model of dynamo action is
currently available \citep{jonesetal2010,char2014}. Direct numerical
simulations aimed at understanding these interactions are restricted
to parameters well away from those pertaining to stellar interiors
(with Reynolds numbers and magnetic Reynolds numbers (\Rm{}) orders of
magnitudes smaller than would be realistic). For this reason, much
attention has been focused on mean-field models of dynamo action
\citep{krause1980}. In this paradigm, only the large-scale flows and
magnetic fields are modelled, with small-scale interactions being
parameterised via transport coefficients such as the $\alpha$-tensor
and the turbulent diffusivity. Although there are many issues with the
mean-field formalism ---~primary among these is whether mean fields
can ever be seen at high \Rm{} or whether the solution is dominated by
the fluctuations~--- these models are of use in describing the
dynamics of mean fields once they have been generated. In particular,
the mean-field equations naturally respect the symmetries of the
underlying rotating spherical system \citep{knobloch1994}, and capture
the nonlinear interactions between magnetic modes of different
symmetries and the underlying large-scale velocity field that is
driving the dynamo.

Mean-field dynamo models have demonstrated that modulation of the
basic cycle may occur through stochastic fluctuations in the
underlying transport coefficients
\citep{schmitt1996,choudhuri2012,hazra2014} or more naturally via
nonlinear interactions inherent in the dynamo equations leading to
chaotic (though deterministic) modulation
\citep{pipin1999,bushbymason2004}. The type of modulation can be
classified according to the key nonlinear interactions that are
primarily responsible \citep{tobias2002}. In the first (Type~1
modulation) magnetic modes of different symmetry (e.g. dipole and
quadrupole modes) interact to produce modulation of the basic cycle,
with significant changes in the symmetry (parity) of solutions. This
behaviour is similar to that seen in the sunspot record over the past
300~years. In the second (imaginatively termed Type~2 modulation) a
magnetic mode with a given symmetry undergoes modulation via
interaction with a large-scale velocity field; here changes in the
amplitude of the basic cycle occur with no significant changes in the
symmetry of solutions. Recently, \citet{weisstobias2016} have argued
from analysis of cosmogenic isotope records that both of these
modulational mechanisms have been at play in the solar dynamo, leading
to `supermodulation' on long time scales. The precise modulational
effects important in the system are sometimes model-dependent. For
this reason, progress can also be made by considering low-order
systems based on symmetry considerations
\citep{knobloch1996,knobloch1998,weiss2011}. These models demonstrate
that the dynamics found in the ad hoc mean-field models is robust and
may be expected in simulations of the full three-dimensional dynamo
system. Symmetry arguments have also proved useful in explaining the
dynamics of dynamo experiments and the geodynamo where the first
bifurcation is stationary \citep{petrelis2009}.

In this paper, we present the results of three-dimensional numerical
solutions of dynamos driven by anelastic convection in a spherical
shell. We do not attempt to model solar convection directly, as this
is well beyond the scope of modern-day computations. Rather, we focus
on the symmetries and nonlinear interactions that lead to modulation
in dynamos, and provide examples of the basic types of modulation and
of supermodulation. These results are important for our understanding
of magnetic field generation via dynamo action, not only in late-type
stars, but also in other astrophysical objects such as planets.

\section{Governing equations}

We consider electrically conducting fluid in a spherical shell
rotating at angular velocity~$\Omega\, \ez$.  The shell is bounded by
two concentric spheres of radius $r_\text{i}$ and $r_\text{o}$ and we
define the shell width~$d=r_\text{o} - r_\text{i}$ and aspect
ratio~$\aspectratio = r_\text{i}/r_\text{o}$. We rely on the LBR
anelastic approximation \citep{braginsky95,lantz99} to model a perfect
gas with kinematic viscosity~$\nu$, turbulent entropy
diffusivity~$\kappa$, specific heat~$c_p$ and magnetic
diffusivity~$\eta$ (all assumed to be constant).  The gravity is given
by ${\vect{g}}=-GM\er/r^2$, where $G$ is the gravitational constant
and $M$ is the central mass.  The equilibrium polytropic solution of
the anelastic system defines the reference state pressure
$\overline{P}=P_c \zeta^{\npol+1}$, density
$\overline{\varrho}=\varrho_c \zeta^\npol$ and temperature
$\overline{T}=T_c \zeta$, with $\zeta=c_0+c_1 d/r$,
$c_0=(2\zeta_0-\aspectratio-1)/(1-\aspectratio)$,
$c_1=(1+\aspectratio)(1-\zeta_o)/(1-\aspectratio)^2$ and
$\zeta_0=(\aspectratio+1)/(\aspectratio\exp(\Nrho /\npol)+1)$. The
constants $P_c$, $\varrho_c$ and $T_c$ are the reference-state
pressure, density and temperature mid-way between the inner and outer
boundaries.  These reference values serve as units for these
variables, whilst length is scaled by~$d$, time by~$d^2/\eta$, entropy
by~$\Delta s$ (the entropy drop across the layer) and magnetic field
by~$\sqrt{\Omega\varrho_c\mu\eta}$, where $\mu$ is the magnetic
permeability. Then, the governing equations are \citep{jones11}
\begin{align}
    \frac{D \vect{v}}{D t} &= \Pm\,
    \bigg[-\frac{1}{\Ekman}\bnabla\frac{P'}{\zeta^n}
      +\frac{\Pm}{\Prandtl}\Ra\frac{s}{r^2} \er -\frac{2}{\Ekman}\,\ez
      \times\vect{v} +
      \vect{F}_\nu+\frac{1}{\Ekman\,\zeta^n}(\bnabla\times\vect{B})\times\vect{B}
      \bigg]\,, \label{mhd1}\\ \frac{\partial\vect{B}}{\partial t} &=
    \bnabla\times(\vect{v}\times\vect{B})+\bnabla^2\vect{B}
    \,,\label{mhd2} \\ \frac{D s}{D t} &=
    \zeta^{-n-1}\frac{\Pm}{\Prandtl}\bnabla\cdot\left(\zeta^{n+1}\,\bnabla
    s\right) + \frac{\Di}{\zeta}\left[\Ekman^{-1}\zeta^{-n}(\bnabla\times
      \vect{B})^2+Q_\nu\right]\,, \label{mhd3}
\end{align}
with the constraints $\bnabla\cdot \left(\zeta^n \vect{v} \right) = 0$
and $\bnabla\cdot \vect{B} = 0$. In the Navier-Stokes
equation~\eqref{mhd1}, $P'$ denotes the pressure perturbation and the
viscous force~$\vect{F}_\nu$ is given by
$\vect{F}_\nu=\zeta^{-n}\bnabla\mathsfbi{S}$, with
$S_{ij}=2\zeta^n\left(e_{ij}-\frac{1}{3}\delta_{ij}\bnabla\cdot
\vect{v}\right)$ and $2e_{ij}=\partial_j v_i+ \partial_i v_j$. The
expressions for the dissipation parameter~\Di{} and the viscous
heating~$Q_\nu$ in~\eqref{mhd3} are $\Di=c_1 \Prandtl/(\Pm \Ra)$ and
$Q_\nu=2\left[e_{ij}e_{ij}-\frac{1}{3}(\bnabla\cdot\vect{v})^2\right]$.
Following \citet{jones11}, we impose stress-free boundary conditions
for the velocity field, and the magnetic field matches a potential
field inside and outside the fluid shell.  The convection is driven by
an imposed entropy difference~$\Delta s$ between the inner and outer
boundaries.  The above system involves seven control parameters: the
Rayleigh number~$\Ra=GMd\Delta s / (\nu\kappa c_p)$, the Ekman
number~$\Ekman =\nu / (\Omega d^2) $, the Prandtl number~$\Prandtl =
\nu / \kappa $, the magnetic Prandtl number~$\Pm = \nu / \eta$,
together with the aspect ratio~\aspectratio{}, the polytropic
index~\npol{} and the number of density scale heights~$\Nrho \equiv
\ln{\left[\overline{\varrho}(r_\text{i})/\overline{\varrho}(r_\text{o})\right]}$.
We set $\Ekman = 10^{-4}$, $\Prandtl=1$, $\Pm=1$, $\aspectratio=0.35$,
$n = 2$ and choose a relatively weak density stratification
$\Nrho=0.5$, to limit the computational time. The critical Rayleigh
number for the linear onset of convection is then $\Rac=3.34 \times
10^5$ \citep[after][]{schrinner2014}.

The anelastic equations are integrated for between 5 and 60 magnetic
diffusion times, which is certainly long enough to establish dynamo
action, utilising the pseudo-spectral code \textsc{parody}
\citep{dormy98}, whose anelastic version \citep{schrinner2014}
reproduces the anelastic dynamo benchmark proposed by
\citet{jones11}. Typical resolutions use 288 points in the radial
direction and a spherical harmonic decomposition truncated at degree
$l_\text{max} \sim 80$ and order $m_\text{max}\sim 60$.  As an
empirical validation of convergence, we ensure for both spectra a
decrease of more than three orders of magnitude over the range of $l$
and $m$. We define the kinetic energy $\Ek = \frac{1}{2}\int\,\zeta^n
\vect{v}^2 \dif V $ and the magnetic energy $\Eb =
\Pm/(2\Ekman)\int\,\vect{B}^2 \dif V$. With our choice of units, a
non-dimensional measure of the velocity amplitude is naturally given
by the magnetic Reynolds number $\Rm = \sqrt{2 \Ek/V}$, $V$ being the
volume of the fluid shell. Crucially for this investigation, which is
concerned with the symmetries of the solutions about the equatorial
plane, we also decompose both the kinetic and magnetic energies
according to their symmetry about the equator (\EkS{}, \EkA{}, \EbS{}
and \EbA{} respectively).  For clarity, we prefer to avoid the terms
dipole and quadrupole families which are also in use to denote the
different parities of the magnetic field; we further adopt the same
definition as \citet{knobloch1998}, according to which
\emph{symmetric} refers to an overall field with \emph{dipole}
symmetry (and vice versa). This choice is consistent with the
properties of pseudo-vectors: the dipole family is invariant under
reflection with respect to the equatorial plane, but the quadrupole
family is not.

\section{Results}

\begin{figure}
  \centering \subfloat[]{%
    \includegraphics[width=0.5\textwidth,trim=80 34 %
      64 50,clip=true]{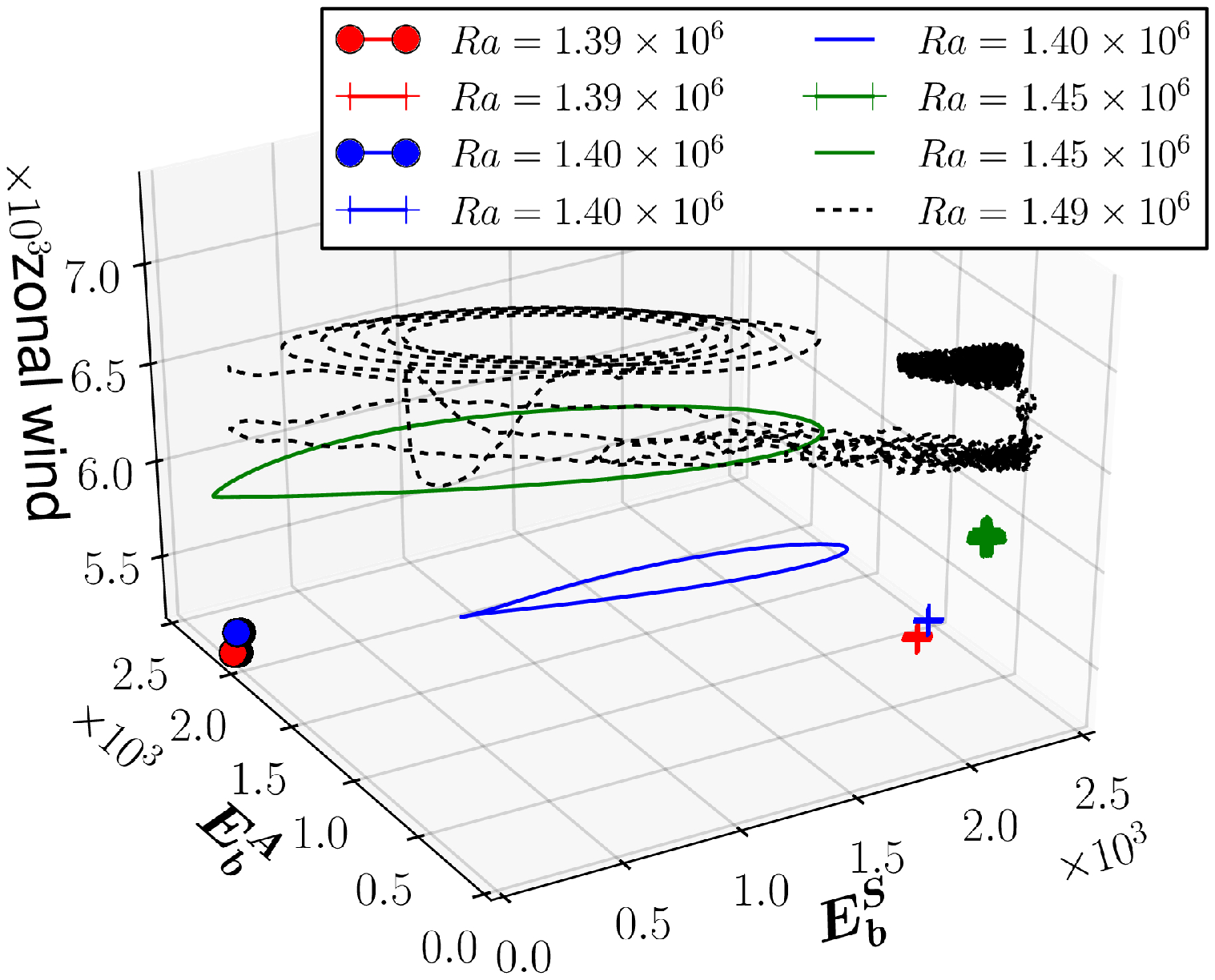}\label{sf:3d}}%
  \hfill%
  \subfloat[$\Ra=1.47\times 10^6$]{%
    \includegraphics[width=0.5\textwidth,trim=20 0 55 34,%
    clip=true]{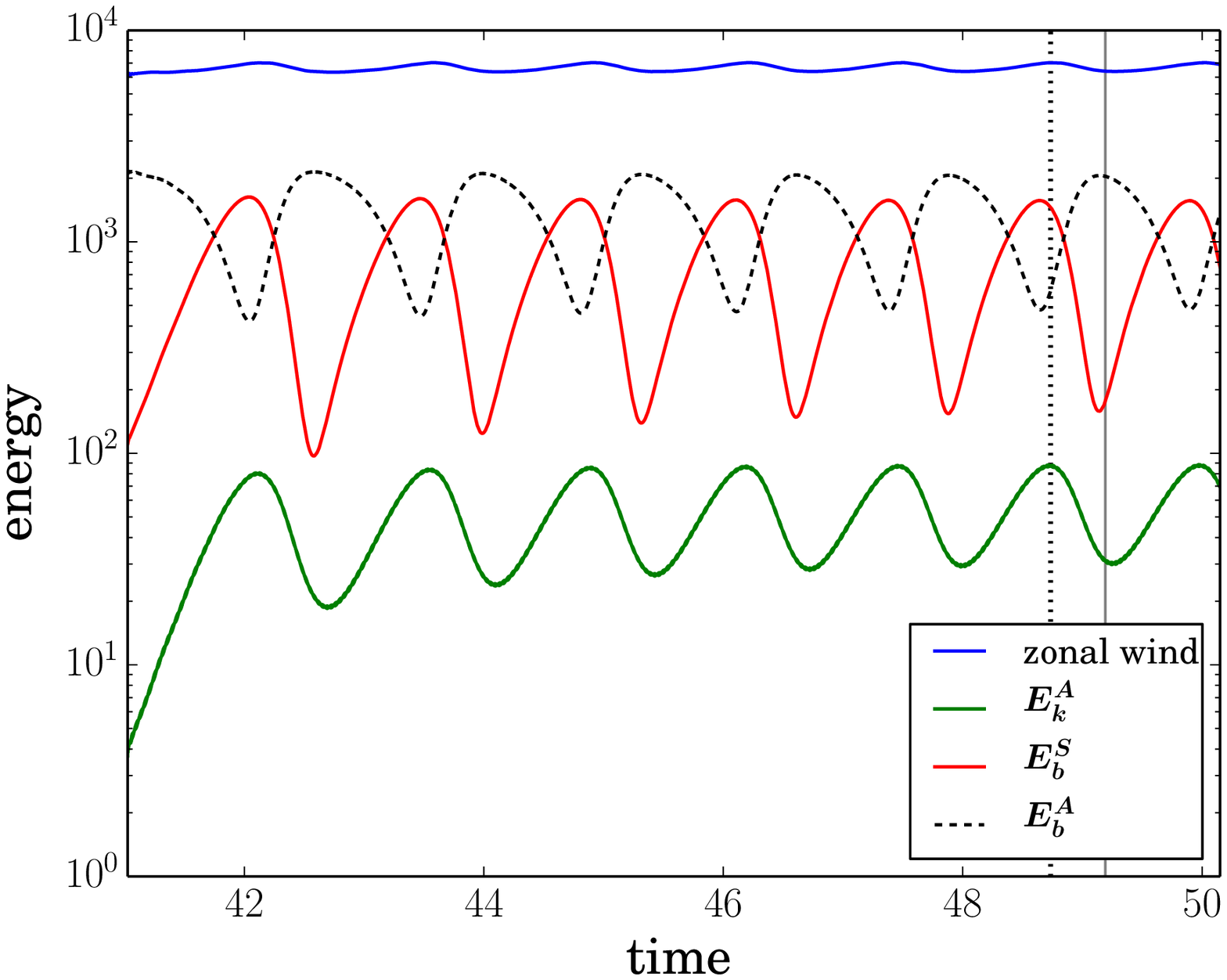}\label{sf:cycle}}%

  \caption{(a): Three-dimensional phase portrait showing the
    projection of the system trajectory onto the space
    $(\EbS,\EbA,\Ez)$. Solid circles denote antisymmetric solutions,
    crosses symmetric solutions. The limit cycle at
    $\Ra=1.40\times10^6$ (solid blue line) has been obtained from a
    similar mixed mode solution after decreasing the value of
    $\Ra$. (b): Energy time series for the limit cycle at
    $\Ra=1.47\times 10^6$.}\label{f:cycles}
\end{figure}

In this paper, we aim to study the symmetry interactions and
low-frequency modulations of the dynamo waves that are characteristics
of the so-called multipolar dynamo branch
\citep{gastine12,schrinner2014}. This branch is the only one that can
be sustained at low magnetic Reynolds number $\Rm \sim 40$
\citep{raynaud2015}. We stress at the outset that the dynamo magnetic
fields we consider here, independent of their symmetry about the
equator, are dominated by their $m=1$ component and note that this
differs from the Sun ---~although the Sun does show a tendency for
active longitudes.  By considering almost Boussinesq models with
$\Nrho=0.1$, \citet{raynaud2014} showed that the non-axisymmetry is
related to the choice of a gravity profile corresponding to a central
mass distribution.  It should be noted that at the low values of \Rm{}
considered here the advective time is comparable with the ohmic
diffusive time (in contrast to stars). These solutions are usually
interpreted in terms of \citet{parker55} waves, in both the Boussinesq
\citep{busse06,schrinner11a,dietrich2013} and anelastic frameworks
\citep{gastine12}, although this interpretation relies on crude
estimates of the $\alpha$-effect via the flow helicity. Following the
methodology of \citet{schrinner12}, we confirm the key role played by
differential rotation in the generation of the toroidal magnetic field
in our sample of models. It is well known that the $\alpha\Omega$
dynamo instability generically sets in as a Hopf bifurcation leading
to oscillatory solutions.  Our aim here is to identify the changes in
the symmetry of the solutions as $\Ra$ is increased with other
parameters held fixed. In practice, we easily distinguish different
branches of solution by restarting from the closest simulations
performed with other parameters; in a few cases, we also tested their
stability by restarting the simulation after killing one or the other
parity of the magnetic field.  At $Ra=1.39 \times 10^6$, the flow does
not break the equatorial symmetry and magnetic modes of different
parity are linearly decoupled.  Depending on the choice of the initial
conditions, we effectively observe a bistability between symmetric and
antisymmetric solutions, illustrated by the red cross and the red dot
in figure~\ref{sf:3d}. In this figure, the trajectory of the system is
projected for different Rayleigh numbers onto the space spanned by the
symmetric and antisymmetric magnetic energies~\EbS{} and \EbA{}, and
the zonal wind energy measured by the axisymmetric toroidal kinetic
energy~\Ez{} ---~a projection introduced by \citet{knobloch1998}. In
spite of a misleading effect of perspective, note that the
contribution of the antisymmetric magnetic field does reduce to a
negligible fraction for the symmetric solutions (crosses), for which
$\EbA/\EbS \leqslant 10^{-2}$. We further stress that the
aforementioned bistability must not be confused with the hysteretic
transition between the dipolar and multipolar branches resulting from
the use of stress-free boundary conditions \citep{schrinner12}.  When
the magnetic field is predominantly antisymmetric, the flow is
characterized by an $m=8$ convection mode; on the other hand, when the
magnetic field is predominantly symmetric, the flow is then
characterized by an $m=9$ convection mode and larger fluctuations of
the kinetic energy. We also note ---~although it is difficult to see
from figure~\ref{sf:3d}~--- that at these parameters the symmetric
mode is quasiperiodic, having undergone a bifurcation from the
periodic state, while the antisymmetric mode is strictly periodic
(taking the form of a dynamo wave).

\begin{figure}
  \centering 
  \subfloat[$t=48.74$]{%
    \includegraphics[width=0.49\textwidth,trim=55 0 55 60,%
      clip=true]{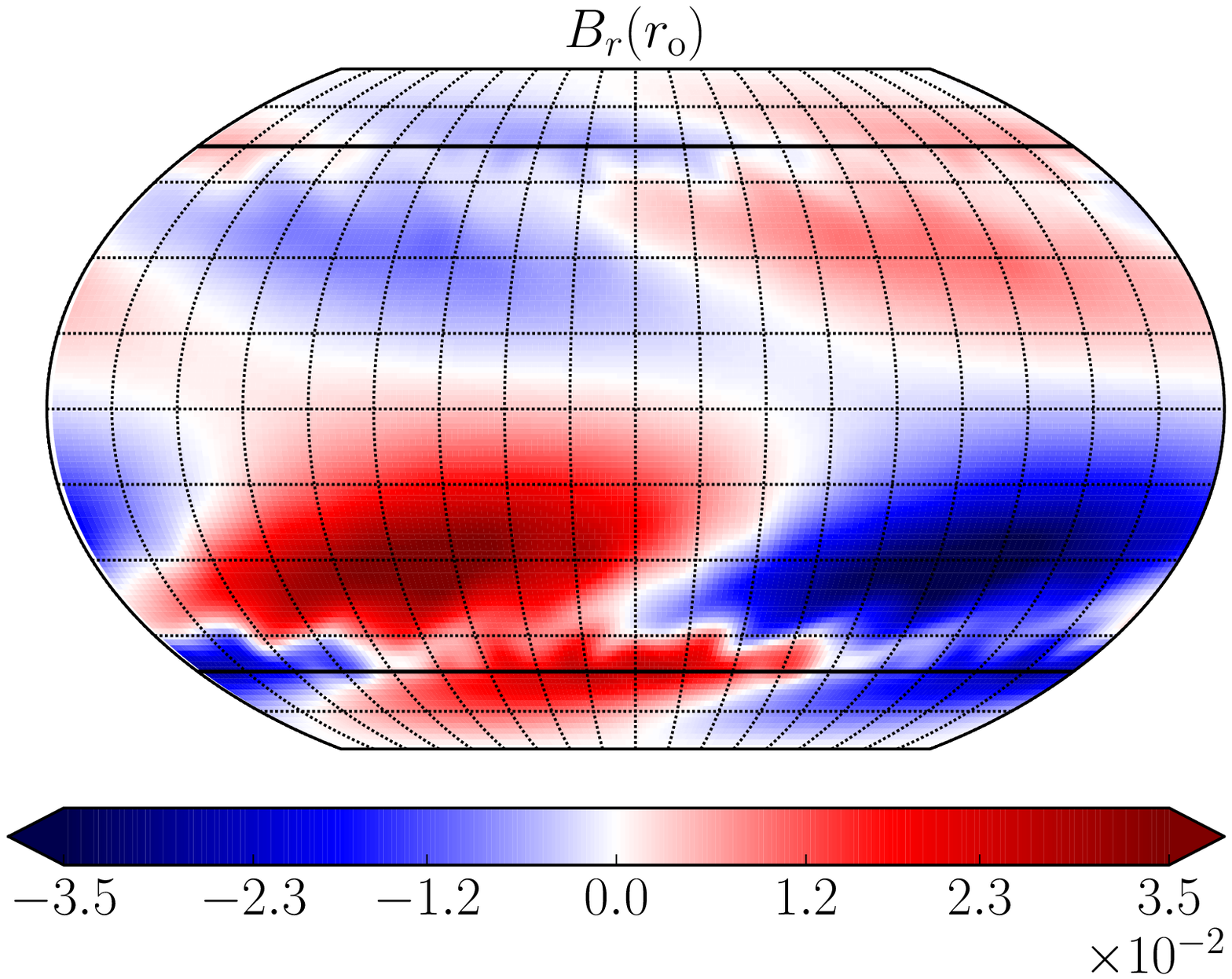}\label{sf:proja}}%
  \hfill%
  \subfloat[$t=49.19$]{%
    \includegraphics[width=0.49\textwidth,trim=55 0 55 60,%
      clip=true]{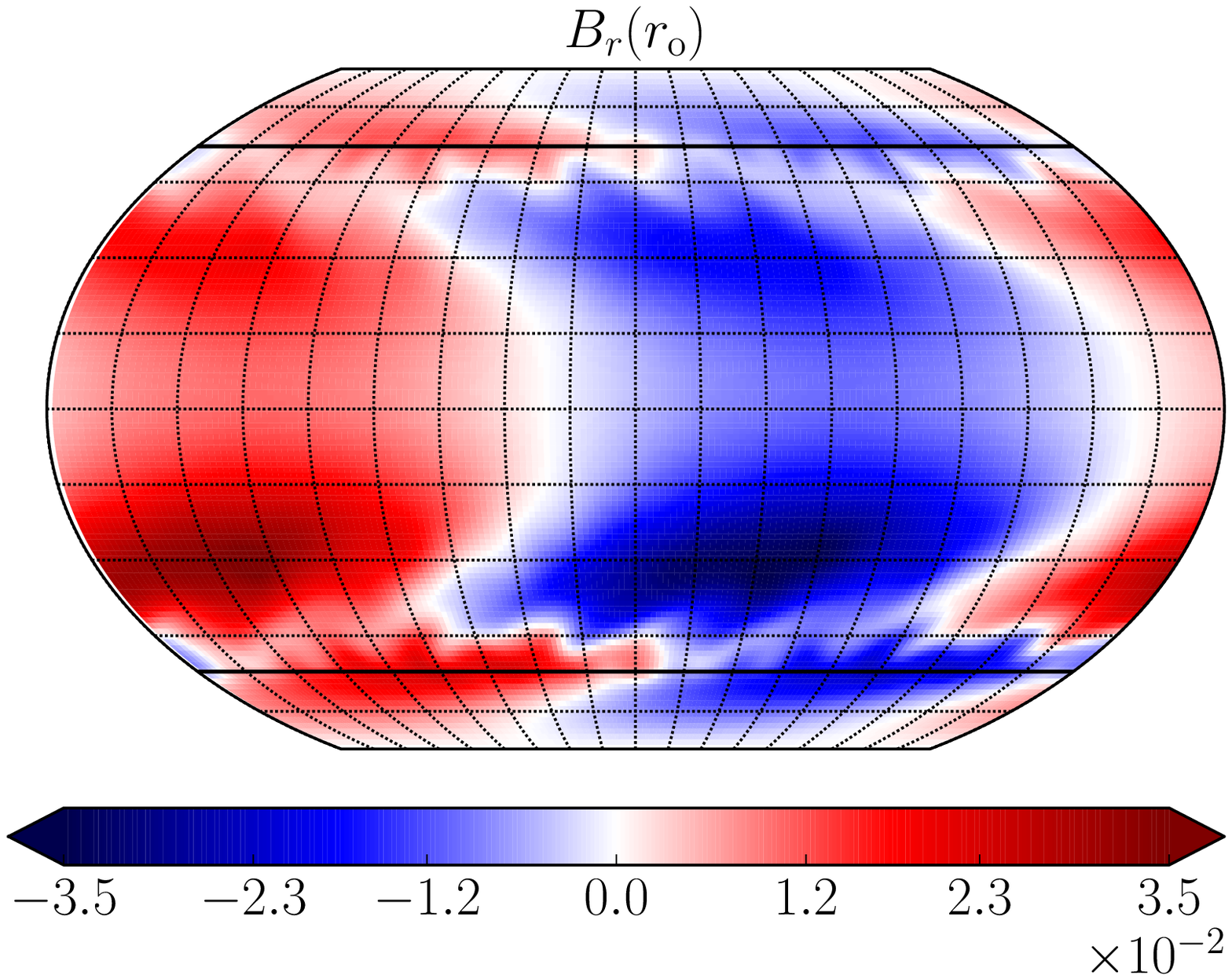}\label{sf:proj}}%
  
  \caption{Snapshots of $B_r$ at the surface of the model for
    $\Ra=1.47\times10^6$ taken at the times highlighted by the dashed
    (a) and solid (b) vertical lines in figure~\ref{sf:cycle}. Note
    that when $\EbS > \EbA$, the magnetic field displays an overall
    dipole symmetry (and conversely).}\label{f:snap}
\end{figure}

Increase of the Rayleigh number from $1.40 \times 10^6$ to $1.45
\times 10^6$ leads to the destabilization of the antisymmetric
solution (blue dot in figure~\ref{sf:3d}) and the discovery of an
asymmetric solution that takes the form of a limit cycle in this phase
space (green solid line); the basic dynamo wave is modulated by change
in the underlying parity of the solution.  This solution coexists with
the symmetric solution (green cross).  An example of the limit cycle
at $Ra=1.47\times 10^6$ is given in figure~\ref{sf:cycle} which shows
the time series of the antisymmetric kinetic energy~\EkA{} (green
solid line), together with those for the symmetric and antisymmetric
magnetic energies~\EbS{} and~\EbA{} (represented by the solid red and
dashed black lines respectively).  This solution is characterized by
weak symmetry breaking of the flow coupling magnetic modes of
different parity. Indeed, we clearly see a periodic exchange of energy
between modes of opposite parity, which could be described as a Type~1
modulation, in reference to the terminology introduced by
\citet{knobloch1998}. Figure~\ref{f:snap} shows projections of the
radial magnetic field at times when the solution is mixed and
antisymmetric. In figure~\ref{sf:proja}, we note that when the
solution is a mixed mode the magnetic field tends to be localized in
one hemisphere \citep{groteBusse2000,gastine12}.  We believe that this
mixed-mode solution is born in a subcritical secondary Hopf
bifurcation from the antisymmetric state.  Evidence for this arises
from the hysteresis that can be identified. As we can see in
figure~\ref{sf:3d}, this state indeed coexists with both the symmetric
and antisymmetric states down to $Ra=1.40 \times 10^6$, below which it
disappears (presumably in a saddle-node bifurcation).  We stress that
what sets the dependence of the period of both the basic cycle and the
modulation of the dynamos for strongly nonlinear solutions is an open
problem and one that is important for understanding stellar activity
\citep{tobias1998,dubchar2013}. In our sample of models, the
modulation period~\Tmod{} is sensitive to the value of the Rayleigh number
but tends toward a constant when it approaches the critical
bifurcation value: for $\Ra \in [1.40\times10^6, 1.43\times10^6]$, we
have $\Tmod \simeq 3.0 \pm 0.2\,$; in contrast, for $\Ra \in
[1.45\times10^6, 1.55\times10^6]$, it seems that $\Tmod \propto
1/\sqrt{\Ra}$, though accurate measurements are compromised by the
fact that we just have 6 data points, 4 of which are only metastable
for $\Ra \geq 1.49\times10^6$.  Figure~\ref{sf:3d} indeed shows that
this mixed-mode limit cycle eventually loses its stability when the
Rayleigh number is increased to $1.49\times 10^6$ (dashed black line)
and the solution develops more sign of spatio-temporal complexity, as
described below.

\begin{figure}
  \centering 
  \subfloat[$\Ra=1.55\times10^6$]{%
    \includegraphics[width=0.33\textwidth,trim=80 34 86 60,%
      clip=true]{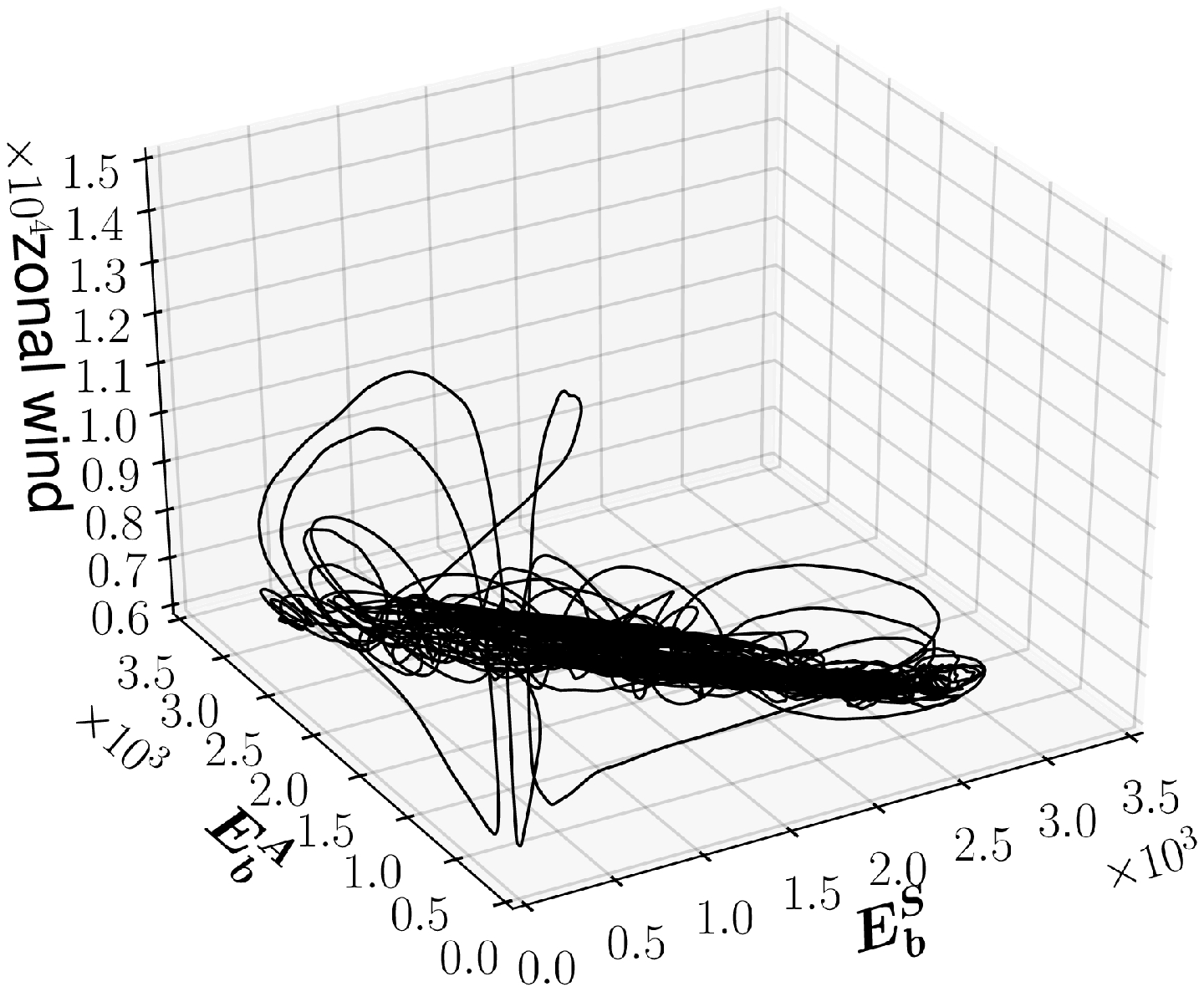}\label{sf:3d155}}%
  \hfill%
  \subfloat[$\Ra=1.65\times10^6$]{%
    \includegraphics[width=0.33\textwidth,trim=80 34 86 60,%
      clip=true]{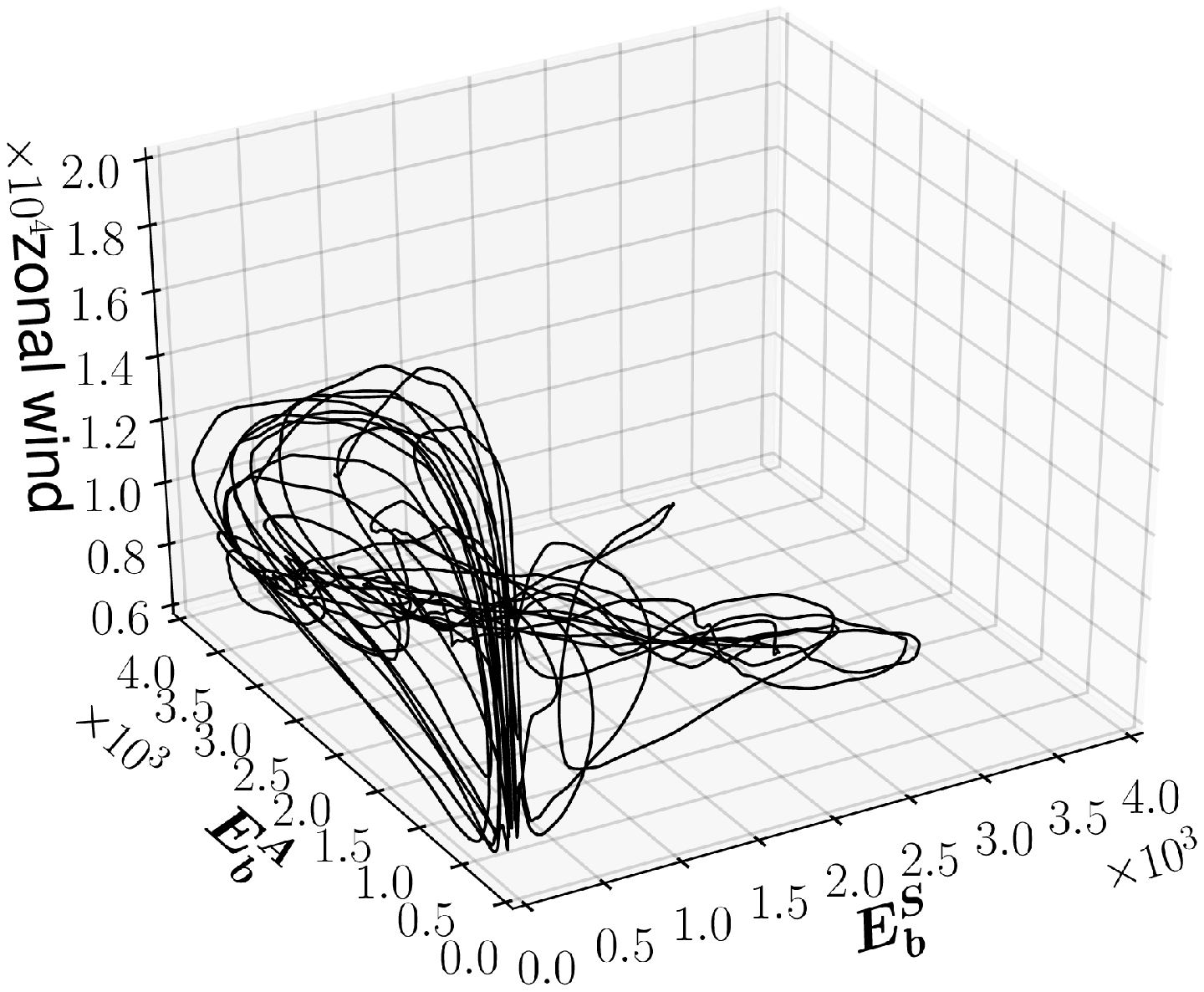}\label{sf:3d165}}%
  \hfill%
  \subfloat[$\Ra=1.85\times10^6$]{%
    \includegraphics[width=0.33\textwidth,trim=80 34 86 60,%
      clip=true]{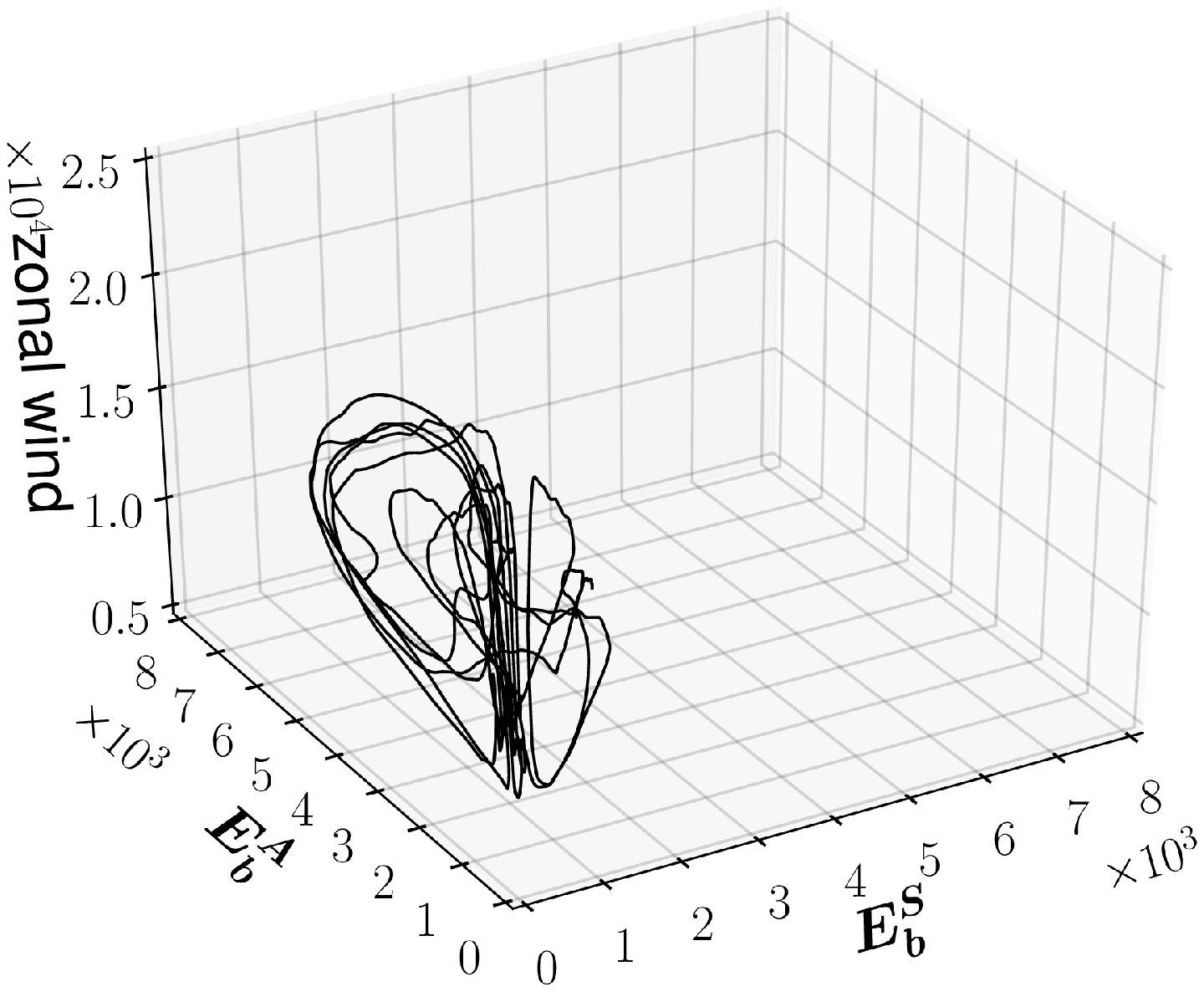}\label{sf:3d185}}%
  
  \caption{Three-dimensional phase portraits showing the projection of
    the system trajectory onto the space
    $(\EbS,\EbA,\Ez)$.}\label{f:type2}
\end{figure}

When the Rayleigh number is further increased, we find that the
dynamics of the magnetic field progressively switches from parity to
amplitude modulations, i.e.\ from Type~1 to Type~2 modulation.  This
transition is particularly clear when comparing the three-dimensional
phase portraits represented for increasing values of the Rayleigh
number in figure~\ref{f:type2}, in which the trajectory of the system
has been smoothed by applying a moving average that removes the basic
dynamo cycle and short-period oscillations.  For $\Ra=1.55\times10^6$
(see figure~\ref{sf:3d155}), the dynamics is mainly governed by the
energy exchange between \EbS{} and \EbA{} (i.e. Type~1 modulation),
and we only distinguish the first signs of the Type~2 modulation
through intermittent decays of the magnetic energy, always followed by
an increase of the zonal wind. In stark contrast, we see in
figure~\ref{sf:3d185} that the system trajectory in the space
$(\EbS,\EbA,\Ez)$ is actually confined near the antisymmetric subspace
(i.e. $\EbS \ll \EbA$) and characterized by the strong amplitude
modulation of the antisymmetric energy by the zonal wind for
$\Ra=1.85\times10^6$.  This is clear Type~2 modulation. Most
interesting, however, is the attractor for $\Ra=1.65 \times 10^6$ in
figure~\ref{sf:3d165}.  This clearly shows the solution exhibiting
both types of modulation; Type~1 modulation where there are no minima
in activity but energy transfer between the modes of different
symmetries and Type~2 modulation where the antisymmetric solution
regularly visits grand minima in activity through interactions with
the zonal wind. The transition between these two types of modulation
has been termed supermodulation and is believed to be prominent in
solar activity records \citep{weisstobias2016}.

\begin{figure}
  \centering%
  \subfloat[$\Ra=1.55\times 10^6$]{%
    \includegraphics[width=0.33\textwidth,trim=20 0 55 26,%
      clip=true]{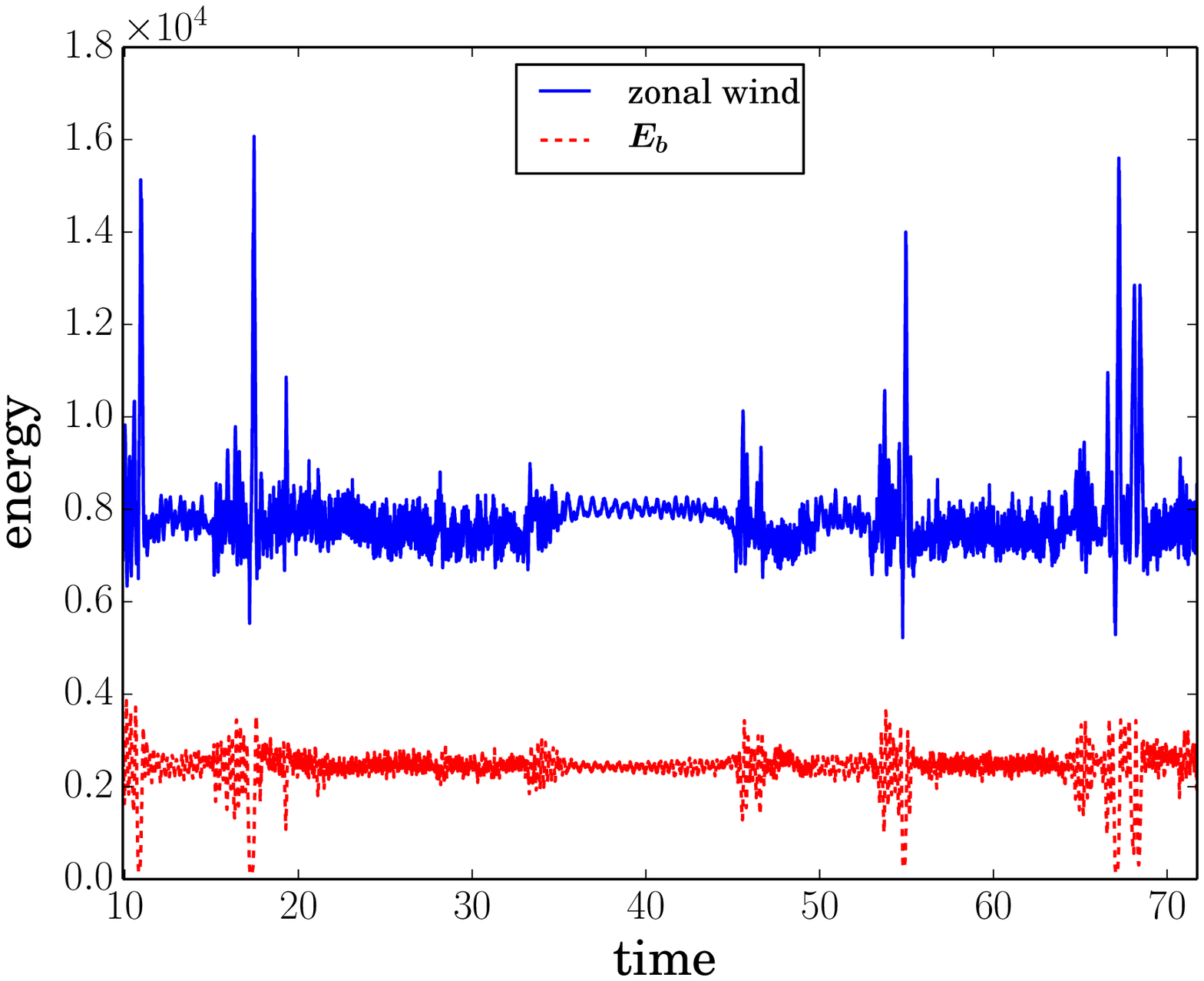}\label{sf:55zw}}%
  \hfill%
  \subfloat[$\Ra=1.65\times 10^6$]{%
    \includegraphics[width=0.33\textwidth,trim=20 0 55 26,%
    clip=true]{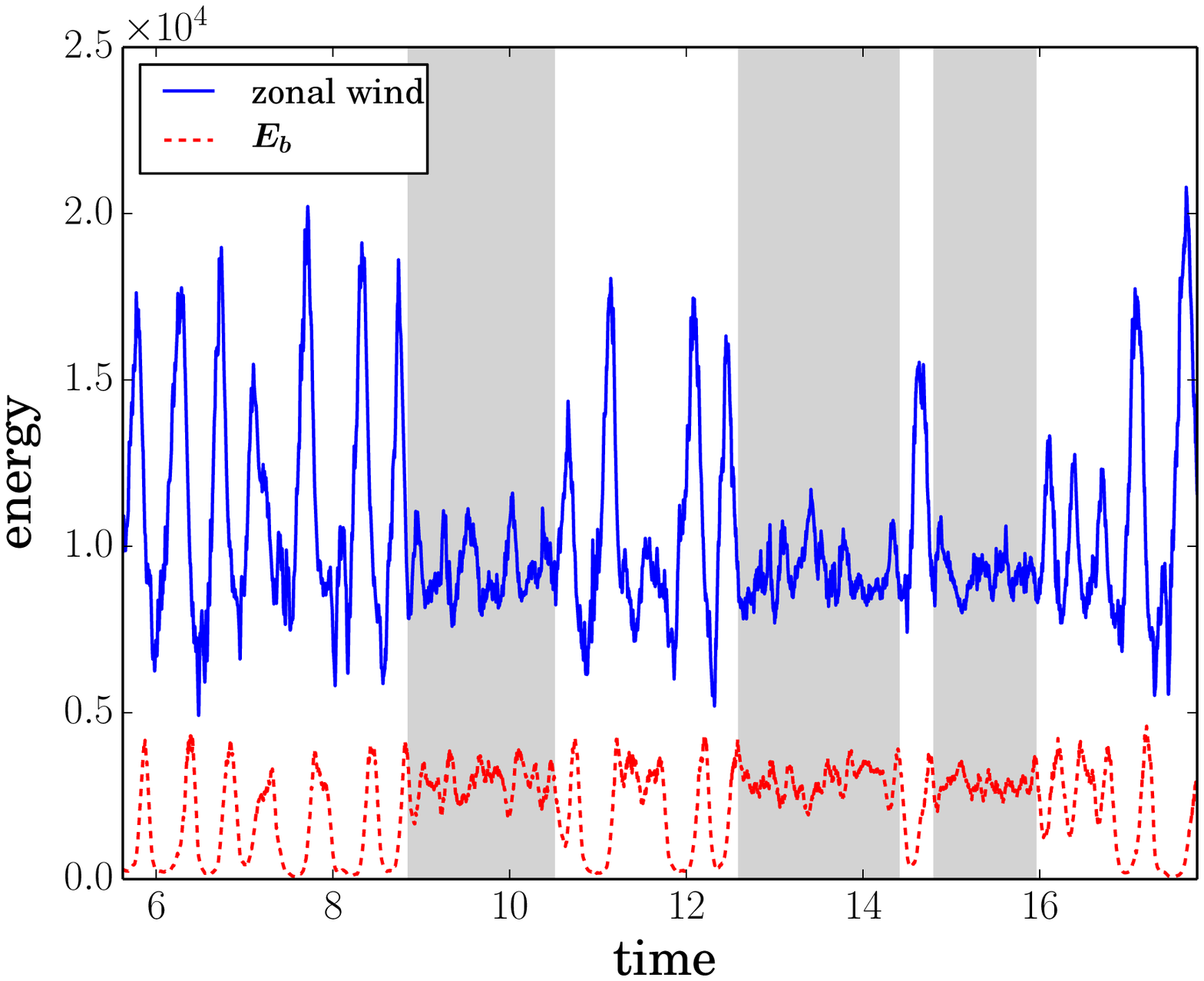}\label{sf:65zw}}%
  \hfill%
  \subfloat[$\Ra=1.85\times 10^6$]{%
    \includegraphics[width=0.33\textwidth,trim=20 0 55 26,%
    clip=true]{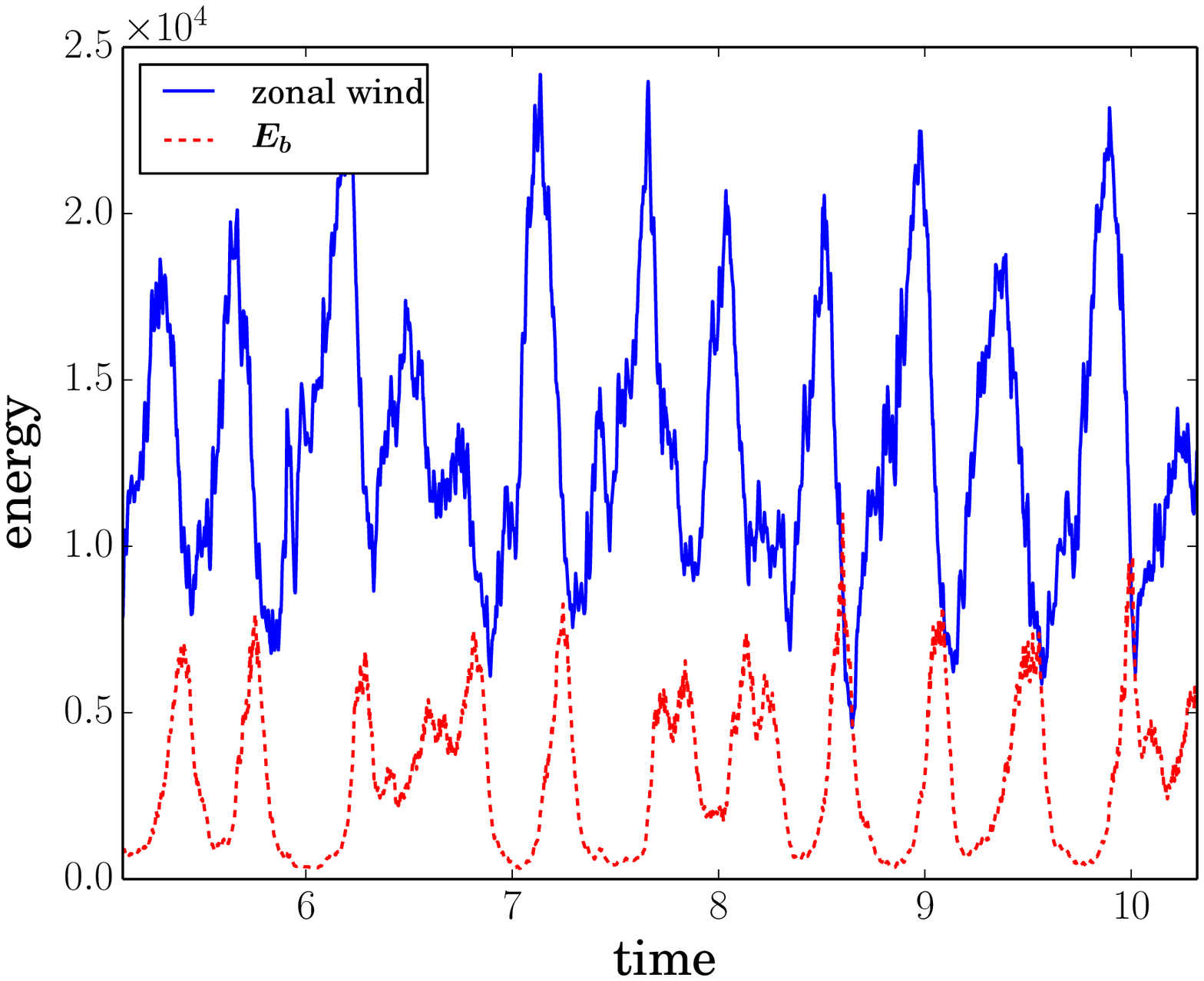}\label{sf:85zw}}%

  \subfloat[$\Ra=1.55\times 10^6$]{%
    \includegraphics[width=0.33\textwidth,trim=20 0 55 26,%
    clip=true]{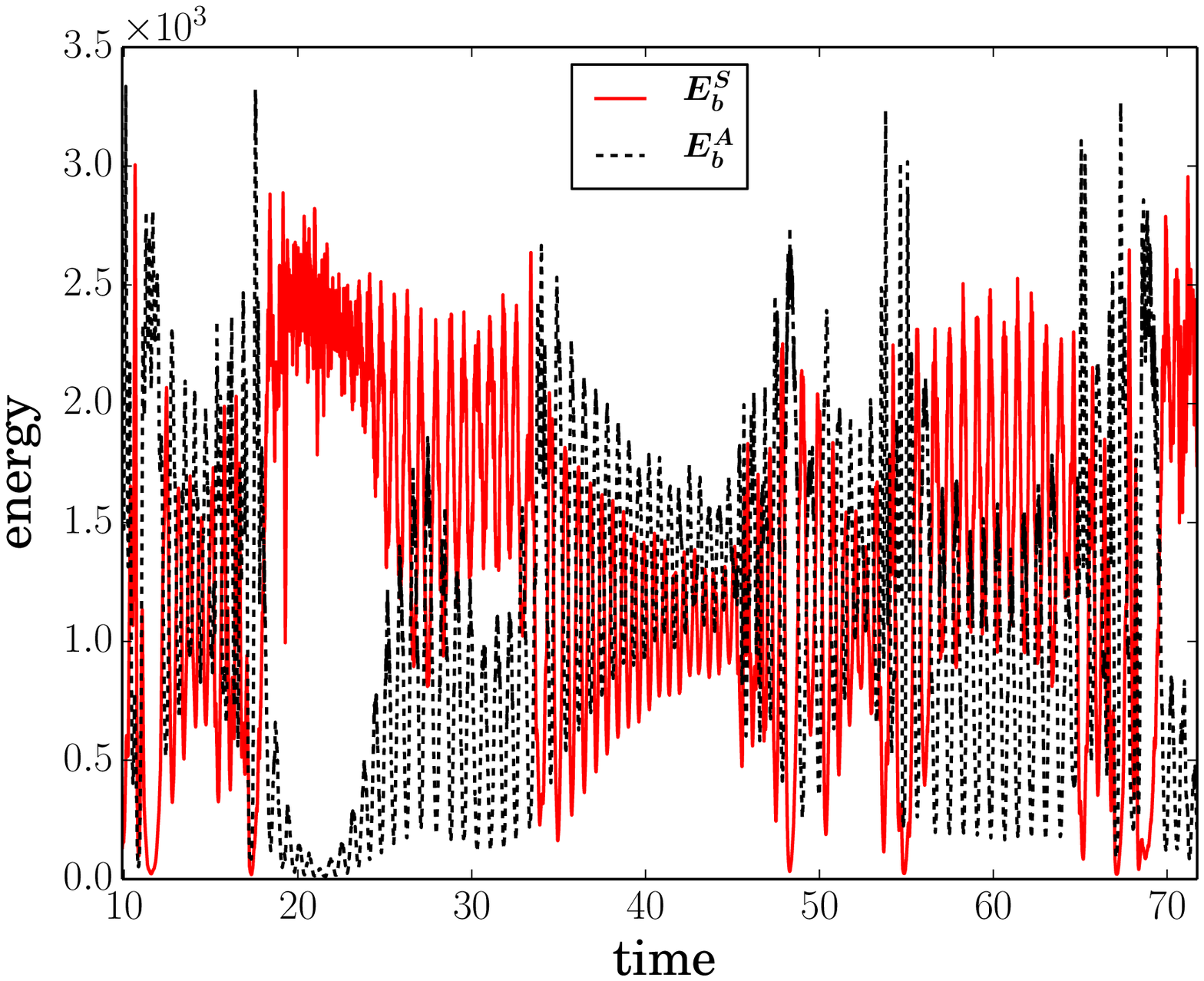}\label{sf:55eb}}%
  \hfill%
  \subfloat[$\Ra=1.65\times 10^6$]{%
    \includegraphics[width=0.33\textwidth,trim=20 0 55 26,%
    clip=true]{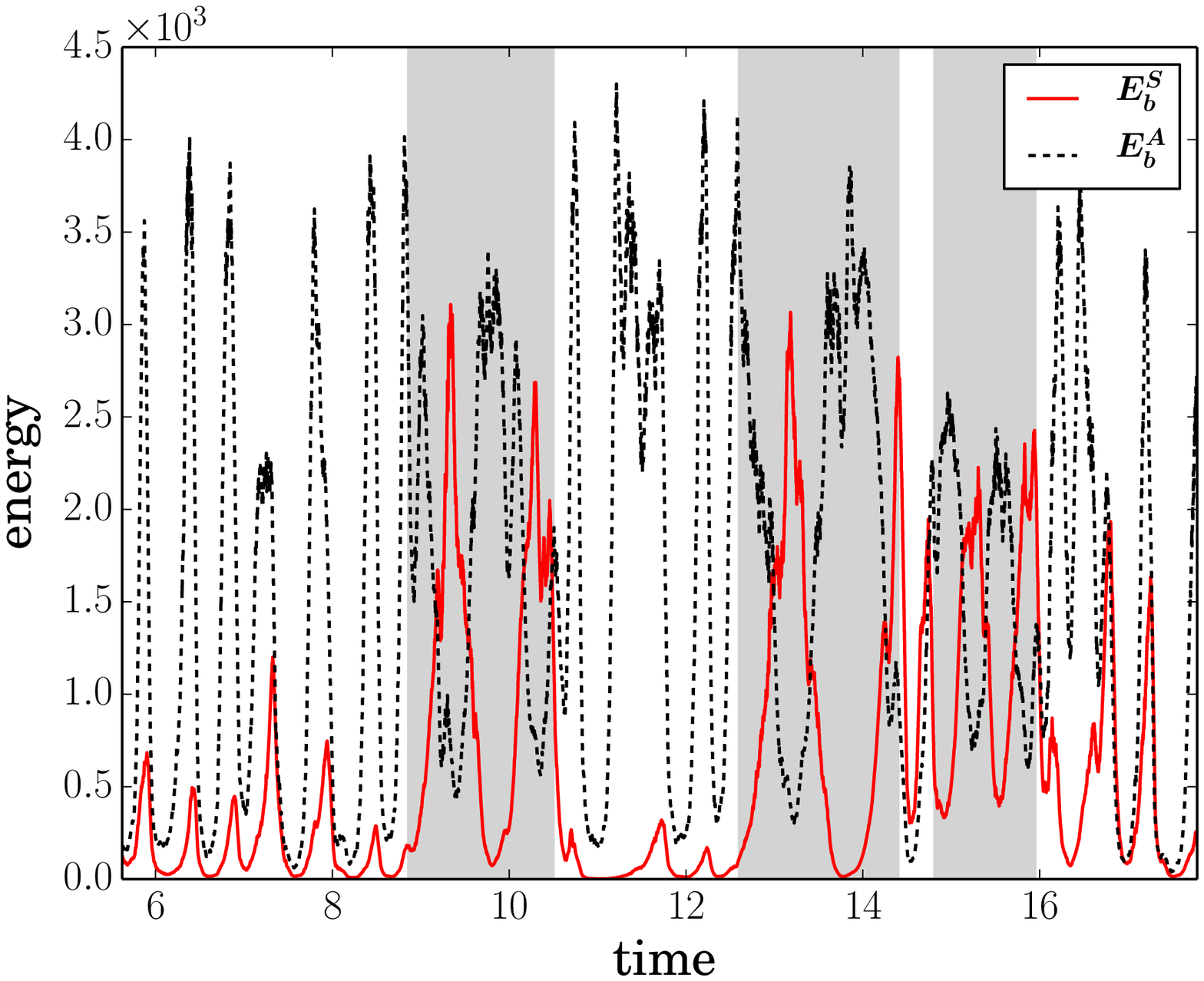}\label{sf:65eb}}%
  \hfill%
  \subfloat[$\Ra=1.85\times 10^6$]{%
    \includegraphics[width=0.33\textwidth,trim=20 0 55 26,%
    clip=true]{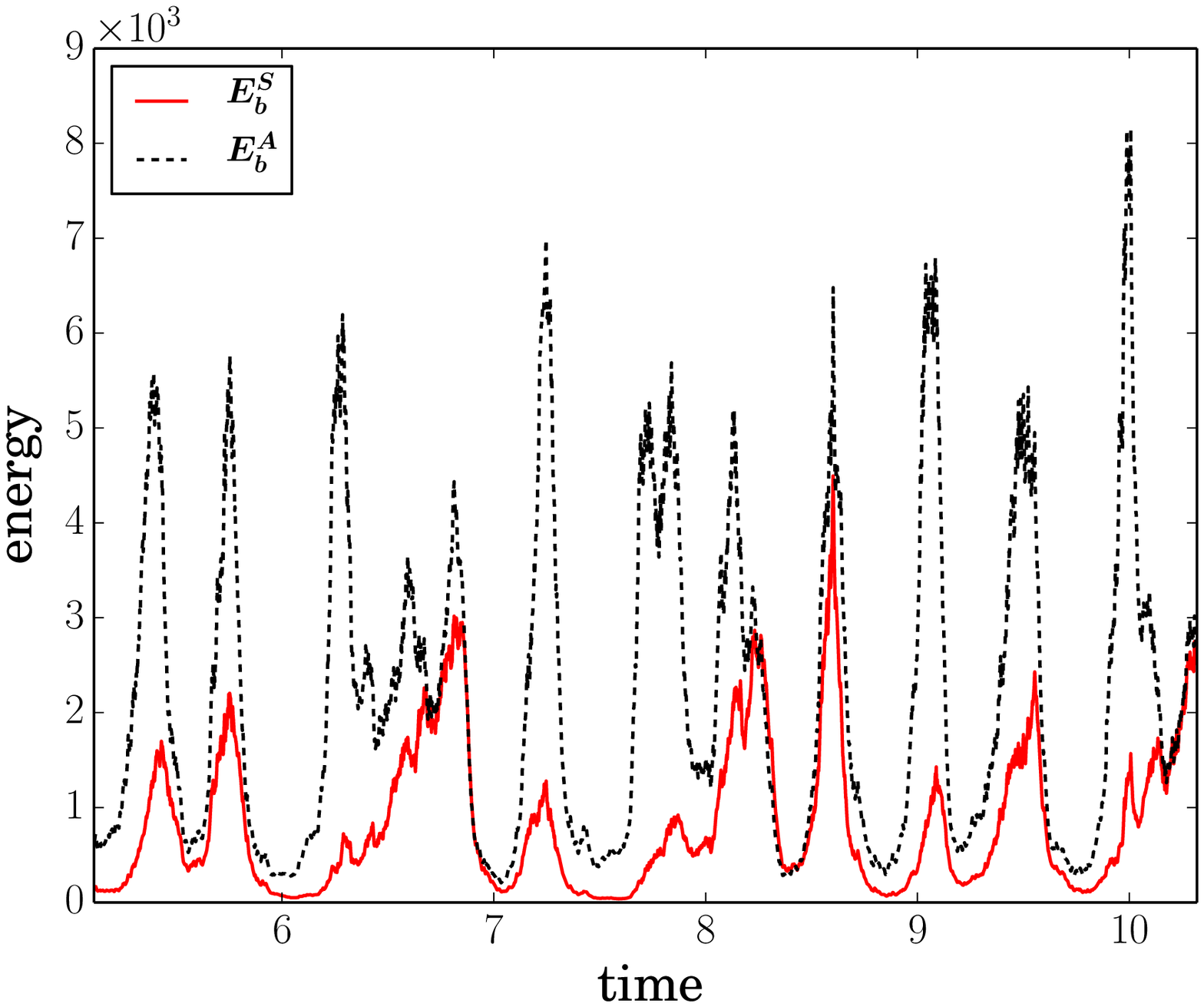}\label{sf:85eb}}%

  \caption{Time series corresponding to the phase portraits in
    figure~\ref{f:type2}. (a--c) Time series of zonal wind energy
    (blue) and total magnetic energy (red). (d--f) Time series of
    symmetric magnetic energy (red) and antisymmetric magnetic energy
    (black). Shaded regions in subfigures~(b) and (e) highlight the
    occurrences of Type~1 modulation.}\label{timeseries_transition}
\end{figure}

The time series corresponding to the trajectories in
figure~\ref{f:type2} are shown in figure~\ref{timeseries_transition}.
For $\Ra=1.55 \times 10^6$, figures~\ref{sf:55zw} and~\ref{sf:55eb}
reveal that the magnetic energy does not exhibit any deep minima
although 4 dips in total energy are visible, and that there are
periods when the symmetric energy is greater than the antisymmetric
energy and periods when they are comparable.  In contrast,
figures~\ref{sf:85zw} and \ref{sf:85eb} show the strong amplitude
modulation of both the zonal wind and the magnetic energy leading to
grand minima observed at $\Ra=1.85\times10^6$. This temporal evolution
is reminiscent of the relaxation oscillations that can be observed in
turbulent hydrodynamic convection \citep{grote2000,christensen02}.  In
that model, the relaxation phenomenon originates from the fact that
the columnar convection feeds the differential rotation through the
action of Reynolds stresses but also tends to be disrupted by the
shear due to differential rotation. This competition between the
convection and the zonal wind is present in our models, since we note,
for instance, that the Nusselt number is always minimum when the zonal
wind reaches its maximum. However, a hydrodynamic simulation performed
at $\Ra=1.75\times10^6$ ---~where the system tends to switch from
supermodulation to pure Type~2 modulation~--- demonstrates that the
flow does not continue to break the equatorial symmetry when the
magnetic field is turned off, and also that there is no amplitude
modulation without the backreaction of the Lorentz force; therefore,
as expected, the magnetic field is playing a key role here. The sudden
growth of the zonal wind results thus from the decrease of the
magnetic field. In general, we observe that the Nusselt number is
higher when the magnetic field is present, which confirms that the
magnetic field promotes the columnar convection and thus the heat
transport by reduction of the zonal wind
\citep{grote2000,yadav2016}. At $Ra=1.85\times10^6$, the antisymmetric
magnetic energy is always larger than that for the symmetric field.  A
closer examination suggests that minima are caused by interactions
with the zonal wind and we report that the typical time scale between
two minima is affected by only 10\% variations of the Prandtl
numbers. It tends to increase at lower \Pm{} (or higher \Prandtl{})
and the modulation even disappears if one of these parameter values is
lowered from 1 to of the order of 0.7 (not shown). More generally,
these observations are compatible with the mean-field dynamo results
of \citet{tobias1996}, who explained the dependence on the magnetic
Prandtl number in terms of the time taken for the zonal velocity to
decay once its energy source has been diminished by the interaction of
magnetic fields with convection.

The supermodulation is shown in figures~\ref{sf:65zw} and
\ref{sf:65eb}, which highlight how the nonlinear solution naturally
transitions between the different types of modulational processes. For
example, in the shaded regions, the solution undergoes changes in
symmetry with no deep minima (Type~1 modulation), while between $t
\approx 6$ and $t \approx 9$ clusters of grand minima are found. From
a mathematical perspective, it is no surprise that a chaotic nonlinear
dynamo solution exhibits such behaviour which has indeed been
predicted \citep{weisstobias2016}.

\begin{figure}
  \centering%
  \subfloat[]{%
    \includegraphics[width=0.33\textwidth, trim=15 0 63 14,%
      clip=true]{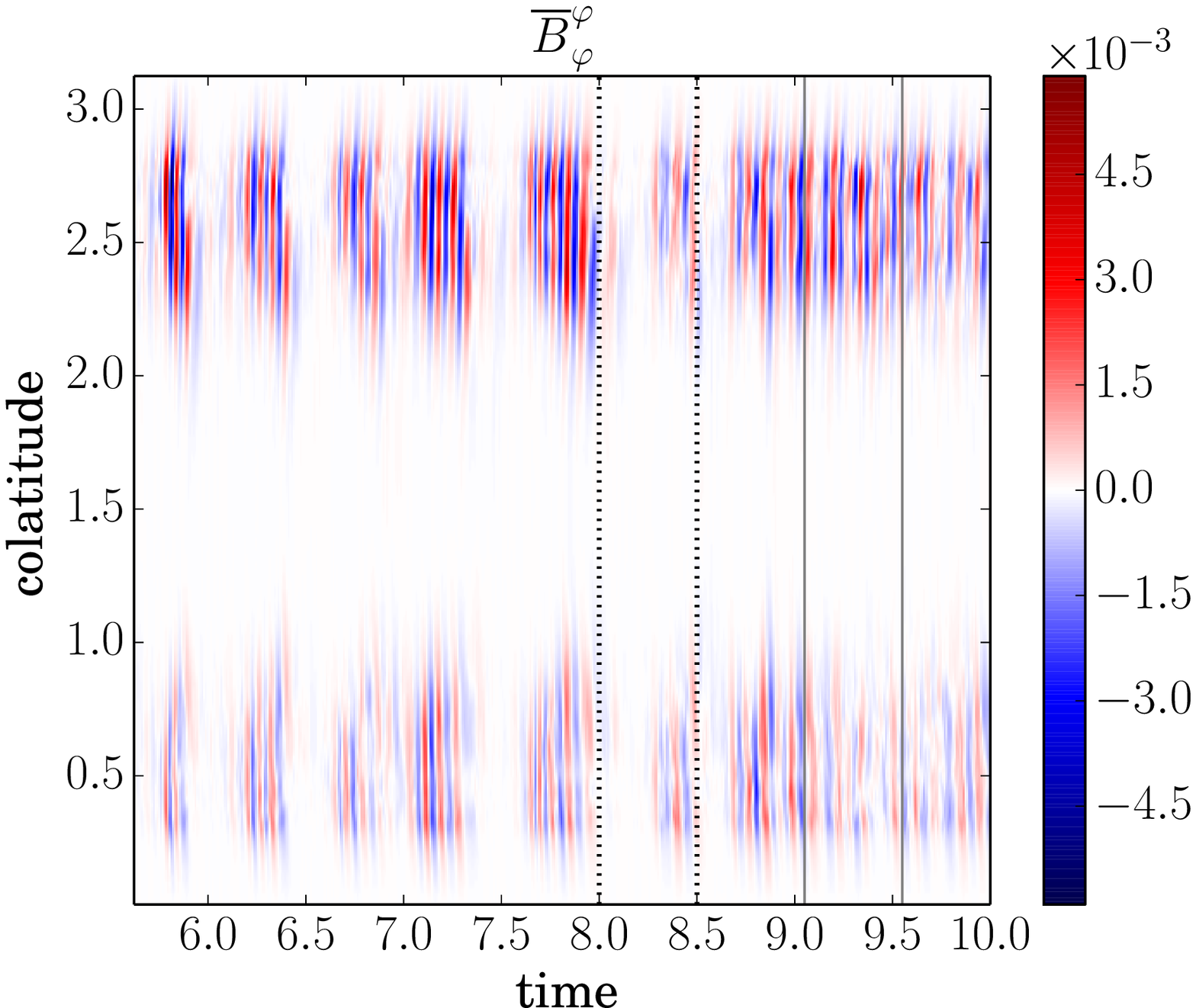}\label{sf:bbp}}%
 \hfill
  \subfloat[]{%
    \includegraphics[width=0.33\textwidth, trim=15 0 60 14,%
    clip=true]{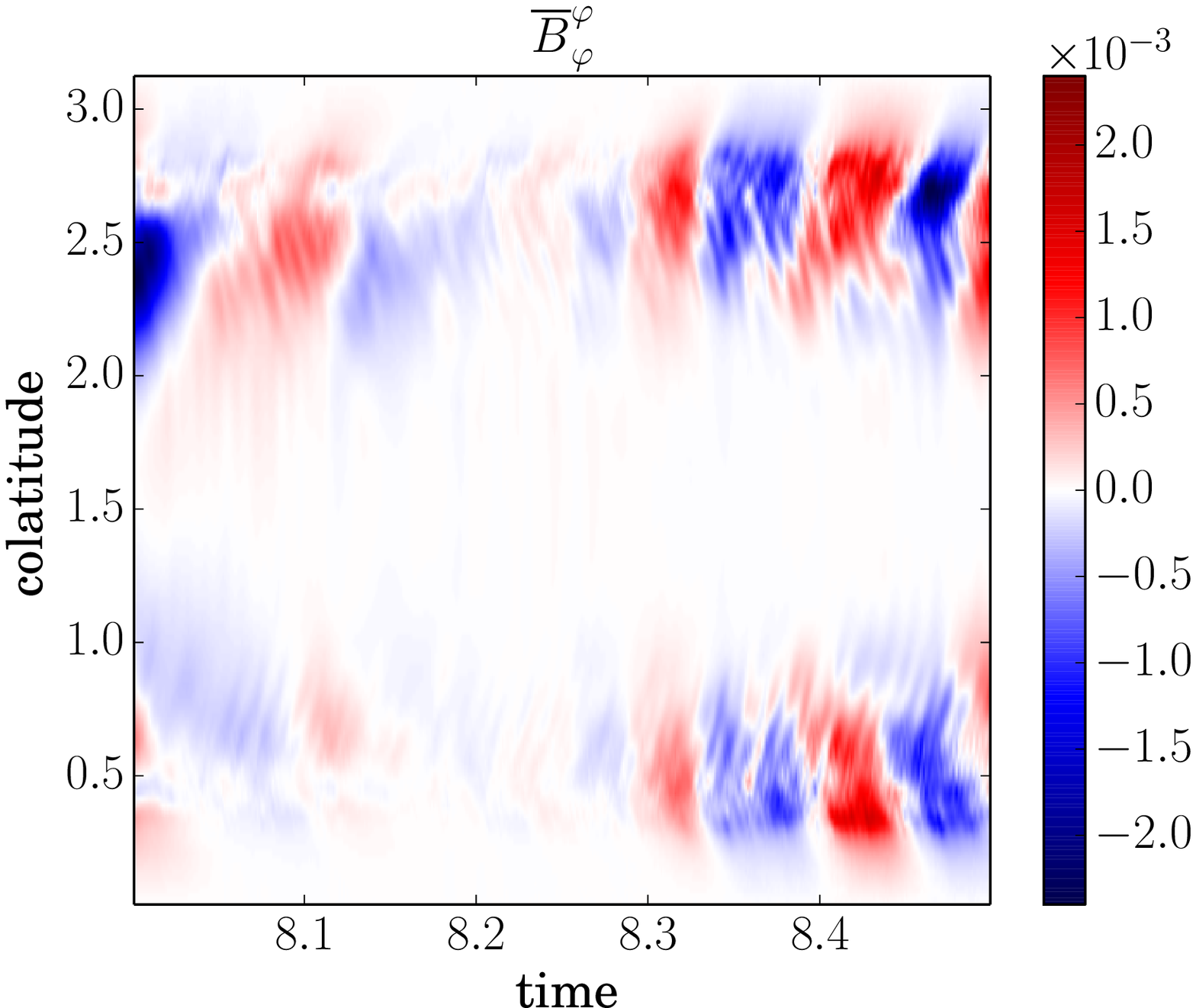}\label{sf:bbp85}}%
 \hfill
  \subfloat[]{%
    \includegraphics[width=0.33\textwidth, trim=15 0 60 14,%
    clip=true]{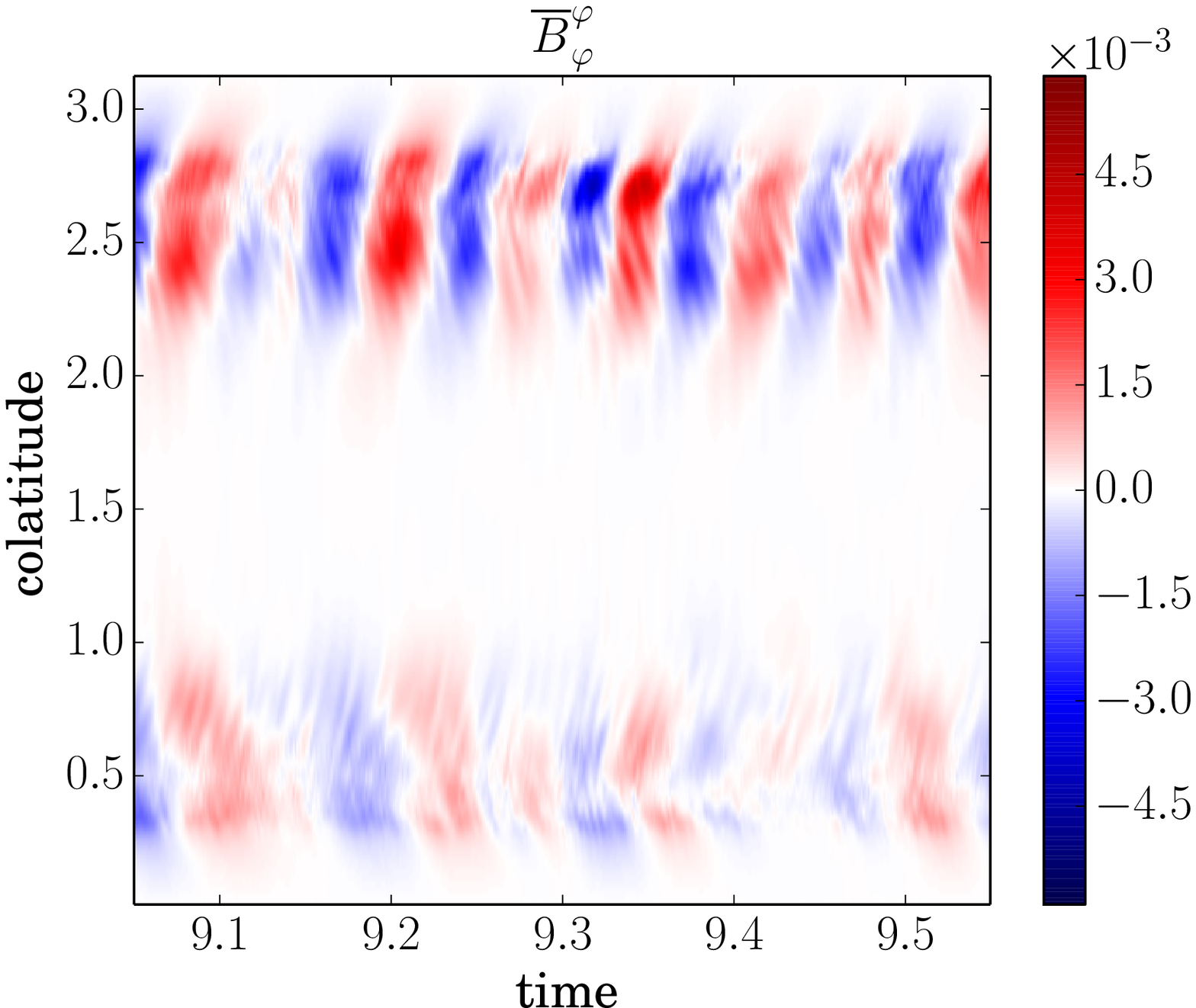}\label{sf:bbr}}%

  \caption{Butterfly diagrams representing the axisymmetric azimuthal
    component of the magnetic field just below the outer surface
    ($r=0.998\,r_\text{o}$) at $\Ra = 1.65\times10^6$. The time
    intervals in subfigures~(b) and (c) correspond to the period
    between the vertical lines in subfigure~(a) ---~dotted for (b) and
    solid for (c).}\label{f:butt}
\end{figure}

Finally, if we examine more closely the evolution of the axisymmetric
magnetic field as a function of colatitude and time, we see in
figure~\ref{f:butt} that both Type~1 and Type~2 modulations affect the
so-called butterfly diagrams in the form of interesting patterns.
These butterfly diagrams correspond to the model with
$\Ra=1.65\times10^6$ whose phase portrait is shown in
figure~\ref{sf:3d165}, and for which supermodulation is present. In
addition to underlining the oscillatory nature of these dynamos, they
demonstrate that both modulational processes occur on time scales that
are not comparable to the period of the dynamo wave, which is in
general of the order of 0.1 magnetic diffusion times in our sample of
models.  Figure~\ref{sf:bbp} illustrates Type~2 modulation for
$t\in[6,8]$, and figure~\ref{sf:bbp85} emphasizes the change in
amplitude of the magnetic field when the system emerges from a grand
minimum. In contrast, the characteristic features of Type~1 modulation
are shown in figure~\ref{sf:bbr}, with the hemispherical magnetic
field undergoing a change of parity at constant amplitude.
Furthermore, the comparison between figures~\ref{sf:bbp85} and
\ref{sf:bbr} (which both cover the same time span) indicates that the
period of the basic cycle is likely to be affected by the superimposed
modulation process, which could be reminiscent of the
\SI{+-30}{\percent}~variability in the duration of the sunspot cycle
\citep{cracken2013}. Although this point deserves further study, we
mention it may not be in contradiction with the simplified Parker wave
dispersion relation, which predicts, for instance, the scaling $\omega
\propto \Ez^{1/4}$ for the frequency of an $\alpha\Omega$ dynamo
\citep{busse06,schrinner11a,gastine12}.  To conclude, we underline
that the magnetic activity appears to be concentrated at high
latitudes, which is probably related to the much smaller aspect ratio
of our models (we recall that we set $\aspectratio=0.35$ whereas the
solar convective zone has an aspect ratio closer to 0.7); this is also
consistent with the fact that the surface magnetic field is
predominantly non-axisymmetric at low latitudes, which results in the
low values displayed by the butterfly diagrams close to the
equator. It should be noted that in all cases the axisymmetric
toroidal field migrates towards the poles, which is reminiscent of the
poleward branch of solar magnetic activity but contrasts with the
equatorward migration of the active latitudes displayed by the solar
butterfly diagrams.

\section{Conclusion}

In this paper we have examined the hydromagnetic interactions between
dynamo modes generated by rotating anelastic convection in a spherical
shell. Motivated by direct and indirect observations of solar magnetic
activity, our primary aim was to investigate the interactions between
modes with different equatorial symmetries.  Mathematically these
dynamos display a dynamical behaviour reminiscent of the results
obtained with (axisymmetric) mean-field models or low-order systems,
with the caveat for the comparison being that the dynamo solutions
presented here are dominated by a non-axisymmetric ($m=1$)
mode. Hemispheric dynamos of the type reported by
\citet{groteBusse2000}, and studied in more detail by
\citet{gallet2009}, have also been found. The present study
demonstrates that this hemispheric configuration is also pertinent to
the understanding of the dynamics of oscillatory dynamos, and thus
could be relevant to explaining the hemispheric magnetic configuration
that has been observed on the Sun at the end of the Maunder minimum
\citep{sokoloff1994, beer1998, knobloch1998}.

We stress again that all current direct numerical simulations of
convective dynamos ---~including those here~--- are far away from what
one can imagine as a ``realistic'' parameter regime. There is,
therefore, the question of the robustness of these results. Of course,
increasing $Ra$ for fixed Ekman number should lead to more disordered
states, gradually breaking all symmetries. What happens after this is
a matter of conjecture/debate. It is possible to argue that for very
high $Ra$ symmetry is re-established on average in the turbulent state
and then similar symmetry-breaking interactions may occur in the
averaged equations. Support for this comes from the finding of such
interactions in mean-field models, which (despite all their drawbacks)
retain the symmetry properties of the underlying system. We note that
symmetry arguments are therefore very powerful and we expect similar
behaviour to be observed in Boussinesq and indeed fully compressible
models.

Our primary result is that we have demonstrated that the interactions
between such modes can lead naturally to a pattern of supermodulation
\citep{arltweiss2014,weisstobias2016} where the system alternates
between modulation with little change of symmetry (with clusters of
deep minima) and modulation that involves significant changes in
symmetry. We believe that this is the first demonstration of such an
interaction between the two types of modulation leading to
supermodulation in the full partial differential equations for
convective dynamos.\\


This study was granted access to the HPC resources of MesoPSL financed
by the R\'{e}gion \^{I}le-de-France and the project Equip@Meso
(reference ANR-10-EQPX-29-01) of the program Investissements d'Avenir
supervised by the Agence Nationale pour la Recherche. Numerical
simulations were also carried out at the TGCC computing center (GENCI
project x2013046698). R.~Raynaud thanks E.~Dormy, C.~Gissinger,
L.~Petitdemange and F.~P\'{e}tr\'{e}lis for various discussions. The
authors thank N.~O.~Weiss for helpful comments.

\bibliographystyle{jfm}
\bibliography{bib/jfm}

\begin{thebibliography}{48}
\expandafter\ifx\csname natexlab\endcsname\relax\def\natexlab#1{#1}\fi
\def\au#1{#1} \def\ed#1{#1} \def\yr#1{#1}\def\at#1{#1}\def\jt#1{\textit{#1}}
  \def\bt#1{#1}\def\bvol#1{\textbf{#1}} \def\vol#1{#1} \def\pg#1{#1}
  \def\publ#1{#1}\def\arxiv#1{#1}\def\org#1{#1}\def\st#1{\textit{#1}}

\bibitem[{Arlt}(2009)]{arlt2009}
{\sc \au{{Arlt}, R.}} \yr{2009}  \at{{The Butterfly Diagram in the Eighteenth
  Century}}.  \jt{\solphys}  \bvol{255},  \pg{143--153},  \arxiv{arXiv:
  0812.2233}.

\bibitem[{Arlt} \& {Weiss}(2014)]{arltweiss2014}
{\sc \au{{Arlt}, R.} \& \au{{Weiss}, N.}} \yr{2014}  \at{{Solar Activity in the
  Past and the Chaotic Behaviour of the Dynamo}}.  \jt{Space Science Review}
  \bvol{186},  \pg{525--533},  \arxiv{arXiv: 1406.7628}.

\bibitem[{Baliunas} {\em et~al.\/}(1998){Baliunas}, {Donahue}, {Soon} \&
  {Henry}]{baliunas1998}
{\sc \au{{Baliunas}, S.~L.}, \au{{Donahue}, R.~A.}, \au{{Soon}, W.} \&
  \au{{Henry}, G.~W.}} \yr{1998}  \at{{Activity Cycles in Lower Main Sequence
  and Post Main Sequence Stars: The HK Project}}.  \jt{Astronomical Society of
  the Pacific Conference Series}  \bvol{154},  \pg{153--172}.

\bibitem[{Beer} {\em et~al.\/}(1998){Beer}, {Tobias} \& {Weiss}]{beer1998}
{\sc \au{{Beer}, J.}, \au{{Tobias}, S.} \& \au{{Weiss}, N.}} \yr{1998}  \at{{An
  Active Sun Throughout the Maunder Minimum}}.  \jt{\solphys}  \bvol{181},
  \pg{237--249}.

\bibitem[{Braginsky} \& {Roberts}(1995)]{braginsky95}
{\sc \au{{Braginsky}, S.~I.} \& \au{{Roberts}, P.~H.}} \yr{1995}
  \at{{Equations governing convection in earth's core and the geodynamo}}.
  \jt{Geophysical and Astrophysical Fluid Dynamics}  \bvol{79},  \pg{1--97}.

\bibitem[{Bushby} \& {Mason}(2004)]{bushbymason2004}
{\sc \au{{Bushby}, P.} \& \au{{Mason}, J.}} \yr{2004}  \at{{Solar dynamo:
  Understanding the solar dynamo}}.  \jt{Astronomy and Geophysics}
  \bvol{45}~(4),  \pg{4.07--4.13}.

\bibitem[{Busse} \& {Simitev}(2006)]{busse06}
{\sc \au{{Busse}, F.~H.} \& \au{{Simitev}, R.~D.}} \yr{2006}  \at{{Parameter
  dependences of convection-driven dynamos in rotating spherical fluid
  shells}}.  \jt{Geophys. Astrophys. Fluid Dyn.}  \bvol{100},  \pg{341--361},
  \arxiv{arXiv: 0904.4293}.

\bibitem[{Charbonneau}(2014)]{char2014}
{\sc \au{{Charbonneau}, P.}} \yr{2014}  \at{{Solar Dynamo Theory}}.  \jt{Annual
  Reviews of Astronomy and Astrophysics}  \bvol{52},  \pg{251--290}.

\bibitem[{Choudhuri} \& {Karak}(2012)]{choudhuri2012}
{\sc \au{{Choudhuri}, A.~R.} \& \au{{Karak}, B.~B.}} \yr{2012}  \at{{Origin of
  Grand Minima in Sunspot Cycles}}.  \jt{Physical Review Letters}
  \bvol{109}~(17),  \pg{171103}.

\bibitem[{Christensen}(2002)]{christensen02}
{\sc \au{{Christensen}, U.~R.}} \yr{2002}  \at{{Zonal flow driven by strongly
  supercritical convection in rotating spherical shells}}.  \jt{Journal of
  Fluid Mechanics}  \bvol{470},  \pg{115--133}.

\bibitem[{Dietrich} {\em et~al.\/}(2013){Dietrich}, {Schmitt} \&
  {Wicht}]{dietrich2013}
{\sc \au{{Dietrich}, W.}, \au{{Schmitt}, D.} \& \au{{Wicht}, J.}} \yr{2013}
  \at{{Hemispherical Parker waves driven by thermal shear in planetary
  dynamos}}.  \jt{EPL (Europhysics Letters)}  \bvol{104},  \pg{49001},
  \arxiv{arXiv: 1402.0343}.

\bibitem[{Dormy} {\em et~al.\/}(1998){Dormy}, {Cardin} \& {Jault}]{dormy98}
{\sc \au{{Dormy}, E.}, \au{{Cardin}, P.} \& \au{{Jault}, D.}} \yr{1998}
  \at{{MHD flow in a slightly differentially rotating spherical shell, with
  conducting inner core, in a dipolar magnetic field}}.  \jt{Earth Planet. Sci.
  Lett.}  \bvol{160},  \pg{15--30}.

\bibitem[{Dub{\'e}} \& {Charbonneau}(2013)]{dubchar2013}
{\sc \au{{Dub{\'e}}, C.} \& \au{{Charbonneau}, P.}} \yr{2013}  \at{{Stellar
  Dynamos and Cycles from Numerical Simulations of Convection}}.
  \jt{Astrophysical Journal}  \bvol{775},  \pg{69}.

\bibitem[{Eddy}(1976)]{eddy1976}
{\sc \au{{Eddy}, J.~A.}} \yr{1976}  \at{{The Maunder Minimum}}.  \jt{Science}
  \bvol{192},  \pg{1189--1202}.

\bibitem[{Gallet} \& {P{\'e}tr{\'e}lis}(2009)]{gallet2009}
{\sc \au{{Gallet}, B.} \& \au{{P{\'e}tr{\'e}lis}, F.}} \yr{2009}  \at{{From
  reversing to hemispherical dynamos}}.  \jt{\pre}  \bvol{80}~(3),
  \pg{035302},  \arxiv{arXiv: 0907.4428}.

\bibitem[{Gastine} {\em et~al.\/}(2012){Gastine}, {Duarte} \&
  {Wicht}]{gastine12}
{\sc \au{{Gastine}, T.}, \au{{Duarte}, L.} \& \au{{Wicht}, J.}} \yr{2012}
  \at{{Dipolar versus multipolar dynamos: the influence of the background
  density stratification}}.  \jt{\aap}  \bvol{546},  \pg{A19},  \arxiv{arXiv:
  1208.6093}.

\bibitem[{Grote} \& {Busse}(2000)]{groteBusse2000}
{\sc \au{{Grote}, E.} \& \au{{Busse}, F.~H.}} \yr{2000}  \at{{Hemispherical
  dynamos generated by convection in rotating spherical shells}}.  \jt{\pre}
  \bvol{62},  \pg{4457}.

\bibitem[{Grote} {\em et~al.\/}(2000){Grote}, {Busse} \& {Tilgner}]{grote2000}
{\sc \au{{Grote}, E.}, \au{{Busse}, F.~H.} \& \au{{Tilgner}, A.}} \yr{2000}
  \at{{Regular and chaotic spherical dynamos}}.  \jt{Physics of the Earth and
  Planetary Interiors}  \bvol{117},  \pg{259--272}.

\bibitem[{Hackman} {\em et~al.\/}(2016){Hackman}, {Lehtinen}, {Ros{\'e}n},
  {Kochukhov} \& {K{\"a}pyl{\"a}}]{hlrk2016}
{\sc \au{{Hackman}, T.}, \au{{Lehtinen}, J.}, \au{{Ros{\'e}n}, L.},
  \au{{Kochukhov}, O.} \& \au{{K{\"a}pyl{\"a}}, M.~J.}} \yr{2016}
  \at{{Zeeman-Doppler imaging of active young solar-type stars}}.  \jt{\aap}
  \bvol{587},  \pg{A28},  \arxiv{arXiv: 1509.02285}.

\bibitem[{Hazra} {\em et~al.\/}(2014){Hazra}, {Passos} \& {Nandy}]{hazra2014}
{\sc \au{{Hazra}, S.}, \au{{Passos}, D.} \& \au{{Nandy}, D.}} \yr{2014}  \at{{A
  Stochastically Forced Time Delay Solar Dynamo Model: Self-consistent Recovery
  from a Maunder-like Grand Minimum Necessitates a Mean-field Alpha Effect}}.
  \jt{\apj}  \bvol{789},  \pg{5},  \arxiv{arXiv: 1307.5751}.

\bibitem[{Jones} {\em et~al.\/}(2011){Jones}, {Boronski}, {Brun}, {Glatzmaier},
  {Gastine}, {Miesch} \& {Wicht}]{jones11}
{\sc \au{{Jones}, C.~A.}, \au{{Boronski}, P.}, \au{{Brun}, A.~S.},
  \au{{Glatzmaier}, G.~A.}, \au{{Gastine}, T.}, \au{{Miesch}, M.~S.} \&
  \au{{Wicht}, J.}} \yr{2011}  \at{{Anelastic convection-driven dynamo
  benchmarks}}.  \jt{\icarus}  \bvol{216},  \pg{120--135}.

\bibitem[{Jones} {\em et~al.\/}(2010){Jones}, {Thompson} \&
  {Tobias}]{jonesetal2010}
{\sc \au{{Jones}, C.~A.}, \au{{Thompson}, M.~J.} \& \au{{Tobias}, S.~M.}}
  \yr{2010}  \at{{The Solar Dynamo}}.  \jt{Space Science Reviews}  \bvol{152},
  \pg{591--616}.

\bibitem[{Knobloch}(1994)]{knobloch1994}
{\sc \au{{Knobloch}, E.}} \yr{1994} {Bifurcations in Rotating Systems}.  \bt{In
  {\em Lectures on Solar and Planetary Dynamos\/} (ed. \ed{M.~R.~E. {Proctor}
  \& A.~D. {Gilbert}})},  \pg{p. 331}.

\bibitem[{Knobloch} \& {Landsberg}(1996)]{knobloch1996}
{\sc \au{{Knobloch}, E.} \& \au{{Landsberg}, A.~S.}} \yr{1996}  \at{{A new
  model of the solar cycle}}.  \jt{\mnras}  \bvol{278},  \pg{294--302}.

\bibitem[{Knobloch} {\em et~al.\/}(1998){Knobloch}, {Tobias} \&
  {Weiss}]{knobloch1998}
{\sc \au{{Knobloch}, E.}, \au{{Tobias}, S.~M.} \& \au{{Weiss}, N.~O.}}
  \yr{1998}  \at{{Modulation and symmetry changes in stellar dynamos}}.
  \jt{\mnras}  \bvol{297},  \pg{1123--1138}.

\bibitem[{Krause} \& {R\"{a}dler}(1980)]{krause1980}
{\sc \au{{Krause}, F.} \& \au{{R\"{a}dler}, K.~H.}} \yr{1980} {\em {Mean-field
  magnetohydrodynamics and dynamo theory}\/}.  \publ{Pergamon Press}.

\bibitem[{Lantz} \& {Fan}(1999)]{lantz99}
{\sc \au{{Lantz}, S.~R.} \& \au{{Fan}, Y.}} \yr{1999}  \at{{Anelastic
  Magnetohydrodynamic Equations for Modeling Solar and Stellar Convection
  Zones}}.  \jt{\apjs}  \bvol{121},  \pg{247--264}.

\bibitem[{McCracken} {\em et~al.\/}(2013){McCracken}, {Beer}, {Steinhilber} \&
  {Abreu}]{cracken2013}
{\sc \au{{McCracken}, K.~G.}, \au{{Beer}, J.}, \au{{Steinhilber}, F.} \&
  \au{{Abreu}, J.}} \yr{2013}  \at{{A Phenomenological Study of the Cosmic Ray
  Variations over the Past 9400 Years, and Their Implications Regarding Solar
  Activity and the Solar Dynamo}}.  \jt{\solphys}  \bvol{286},  \pg{609--627}.

\bibitem[{Moffatt}(1978)]{moffattHK}
{\sc \au{{Moffatt}, H.~K.}} \yr{1978} {\em {Magnetic Field Generation in
  Electrically Conducting Fluids}\/}.  \publ{Cambridge University Press}.

\bibitem[{Ol{\'a}h} {\em et~al.\/}(2009){Ol{\'a}h}, {Koll{\'a}th}, {Granzer},
  {Strassmeier}, {Lanza}, {J{\"a}rvinen}, {Korhonen}, {Baliunas}, {Soon},
  {Messina} \& {Cutispoto}]{olah2009}
{\sc \au{{Ol{\'a}h}, K.}, \au{{Koll{\'a}th}, Z.}, \au{{Granzer}, T.},
  \au{{Strassmeier}, K.~G.}, \au{{Lanza}, A.~F.}, \au{{J{\"a}rvinen}, S.},
  \au{{Korhonen}, H.}, \au{{Baliunas}, S.~L.}, \au{{Soon}, W.}, \au{{Messina},
  S.} \& \au{{Cutispoto}, G.}} \yr{2009}  \at{{Multiple and changing cycles of
  active stars. II. Results}}.  \jt{\aap}  \bvol{501},  \pg{703--713}.

\bibitem[{Parker}(1955)]{parker55}
{\sc \au{{Parker}, E.~N.}} \yr{1955}  \at{{Hydromagnetic Dynamo Models.}}
  \jt{\apj}  \bvol{122},  \pg{293--+}.

\bibitem[{P{\'e}tr{\'e}lis} {\em et~al.\/}(2009){P{\'e}tr{\'e}lis}, {Fauve},
  {Dormy} \& {Valet}]{petrelis2009}
{\sc \au{{P{\'e}tr{\'e}lis}, F.}, \au{{Fauve}, S.}, \au{{Dormy}, E.} \&
  \au{{Valet}, J.-P.}} \yr{2009}  \at{{Simple Mechanism for Reversals of
  Earth's Magnetic Field}}.  \jt{Physical Review Letters}  \bvol{102}~(14),
  \pg{144503},  \arxiv{arXiv: 0806.3756}.

\bibitem[{Pipin}(1999)]{pipin1999}
{\sc \au{{Pipin}, V.~V.}} \yr{1999}  \at{{The Gleissberg cycle by a nonlinear
  $\alpha \Lambda$ dynamo}}.  \jt{Astronomy and Astrophysics}  \bvol{346},
  \pg{295--302}.

\bibitem[{Raynaud} {\em et~al.\/}(2014){Raynaud}, {Petitdemange} \&
  {Dormy}]{raynaud2014}
{\sc \au{{Raynaud}, R.}, \au{{Petitdemange}, L.} \& \au{{Dormy}, E.}} \yr{2014}
   \at{{Influence of the mass distribution on the magnetic field topology}}.
  \jt{\aap}  \bvol{567},  \pg{A107},  \arxiv{arXiv: 1406.4743}.

\bibitem[{Raynaud} {\em et~al.\/}(2015){Raynaud}, {Petitdemange} \&
  {Dormy}]{raynaud2015}
{\sc \au{{Raynaud}, R.}, \au{{Petitdemange}, L.} \& \au{{Dormy}, E.}} \yr{2015}
   \at{{Dipolar dynamos in stratified systems}}.  \jt{\mnras}  \bvol{448},
  \pg{2055--2065},  \arxiv{arXiv: 1503.00165}.

\bibitem[{Schmitt} {\em et~al.\/}(1996){Schmitt}, {Schuessler} \&
  {Ferriz-Mas}]{schmitt1996}
{\sc \au{{Schmitt}, D.}, \au{{Schuessler}, M.} \& \au{{Ferriz-Mas}, A.}}
  \yr{1996}  \at{{Intermittent solar activity by an on-off dynamo.}}  \jt{\aap}
   \bvol{311},  \pg{L1--L4}.

\bibitem[{Schrinner} {\em et~al.\/}(2011){Schrinner}, {Petitdemange} \&
  {Dormy}]{schrinner11a}
{\sc \au{{Schrinner}, M.}, \au{{Petitdemange}, L.} \& \au{{Dormy}, E.}}
  \yr{2011}  \at{{Oscillatory dynamos and their induction mechanisms}}.
  \jt{\aap}  \bvol{530},  \pg{A140},  \arxiv{arXiv: 1101.1837}.

\bibitem[{Schrinner} {\em et~al.\/}(2012){Schrinner}, {Petitdemange} \&
  {Dormy}]{schrinner12}
{\sc \au{{Schrinner}, M.}, \au{{Petitdemange}, L.} \& \au{{Dormy}, E.}}
  \yr{2012}  \at{{Dipole Collapse and Dynamo Waves in Global Direct Numerical
  Simulations}}.  \jt{\apj}  \bvol{752},  \pg{121}.

\bibitem[{Schrinner} {\em et~al.\/}(2014){Schrinner}, {Petitdemange}, {Raynaud}
  \& {Dormy}]{schrinner2014}
{\sc \au{{Schrinner}, M.}, \au{{Petitdemange}, L.}, \au{{Raynaud}, R.} \&
  \au{{Dormy}, E.}} \yr{2014}  \at{{Topology and field strength in spherical,
  anelastic dynamo simulations}}.  \jt{\aap}  \bvol{564},  \pg{A78}.

\bibitem[{Sokoloff} \& {Nesme-Ribes}(1994)]{sokoloff1994}
{\sc \au{{Sokoloff}, D.} \& \au{{Nesme-Ribes}, E.}} \yr{1994}  \at{{The Maunder
  minimum: A mixed-parity dynamo mode~?}}  \jt{\aap}  \bvol{288},
  \pg{293--298}.

\bibitem[Tobias(2002)]{tobias2002}
{\sc \au{Tobias, S.M.}} \yr{2002}  \at{Modulation of solar and stellar
  dynamos}.  \jt{Astronomische Nachrichten}  \bvol{323}~(3-4),  \pg{417--423}.

\bibitem[{Tobias}(1996)]{tobias1996}
{\sc \au{{Tobias}, S.~M.}} \yr{1996}  \at{{Grand minima in nonlinear dynamos}}.
   \jt{\aap}  \bvol{307},  \pg{L21}.

\bibitem[{Tobias}(1998)]{tobias1998}
{\sc \au{{Tobias}, S.~M.}} \yr{1998}  \at{{Relating stellar cycle periods to
  dynamo calculations}}.  \jt{Monthly Notices of the Royal Astronomical
  Society}  \bvol{296},  \pg{653--661}.

\bibitem[{Usoskin}(2013)]{usoskin2013}
{\sc \au{{Usoskin}, I.~G.}} \yr{2013}  \at{{A History of Solar Activity over
  Millennia}}.  \jt{Living Reviews in Solar Physics}  \bvol{10}.

\bibitem[{Usoskin} {\em et~al.\/}(2015){Usoskin}, {Arlt}, {Asvestari},
  {Hawkins}, {K{\"a}pyl{\"a}}, {Kovaltsov}, {Krivova}, {Lockwood}, {Mursula},
  {O'Reilly}, {Owens}, {Scott}, {Sokoloff}, {Solanki}, {Soon} \&
  {Vaquero}]{usoskin2015}
{\sc \au{{Usoskin}, I.~G.}, \au{{Arlt}, R.}, \au{{Asvestari}, E.},
  \au{{Hawkins}, E.}, \au{{K{\"a}pyl{\"a}}, M.}, \au{{Kovaltsov}, G.~A.},
  \au{{Krivova}, N.}, \au{{Lockwood}, M.}, \au{{Mursula}, K.}, \au{{O'Reilly},
  J.}, \au{{Owens}, M.}, \au{{Scott}, C.~J.}, \au{{Sokoloff}, D.~D.},
  \au{{Solanki}, S.~K.}, \au{{Soon}, W.} \& \au{{Vaquero}, J.~M.}} \yr{2015}
  \at{{The Maunder minimum (1645-1715) was indeed a grand minimum: A
  reassessment of multiple datasets}}.  \jt{\aap}  \bvol{581},  \pg{A95},
  \arxiv{arXiv: 1507.05191}.

\bibitem[{Weiss}(2011)]{weiss2011}
{\sc \au{{Weiss}, N.~O.}} \yr{2011}  \at{{Chaotic behaviour in low-order models
  of planetary and stellar dynamos}}.  \jt{Geophysical and Astrophysical Fluid
  Dynamics}  \bvol{105},  \pg{256--272}.

\bibitem[{Weiss} \& {Tobias}(2016)]{weisstobias2016}
{\sc \au{{Weiss}, N.~O.} \& \au{{Tobias}, S.~M.}} \yr{2016}
  \at{{Supermodulation of the Sun's magnetic activity: the effects of symmetry
  changes}}.  \jt{\mnras}  \bvol{456},  \pg{2654--2661}.

\bibitem[{Yadav} {\em et~al.\/}(2016){Yadav}, {Gastine}, {Christensen},
  {Duarte} \& {Reiners}]{yadav2016}
{\sc \au{{Yadav}, R.~K.}, \au{{Gastine}, T.}, \au{{Christensen}, U.~R.},
  \au{{Duarte}, L.~D.~V.} \& \au{{Reiners}, A.}} \yr{2016}  \at{{Effect of
  shear and magnetic field on the heat-transfer efficiency of convection in
  rotating spherical shells}}.  \jt{Geophysical Journal International}
  \bvol{204},  \pg{1120--1133},  \arxiv{arXiv: 1507.03649}.

\end{thebibliography}

\end{document}